\documentclass{emulateapj}
\usepackage{natbib}

\newcommand\etal{et al.}
\newcommand\ie{i.e.}
\newcommand\eg{e.g.}

\newcommand\kms{\ifmmode{\rm km\ s^{-1}}\else$\rm km\ s^{-1}$\fi}
\newcommand\mydeg{$^\circ$}
\def\eps@scaling{1.0}
\newcommand\plotthree[3]{{
 \typeout{Plotthree included the files #1 #2 #3}
 \centering
 \leavevmode
 \columnwidth=.33\columnwidth
 \includegraphics[width={\eps@scaling\columnwidth}]{#1}
 \includegraphics[width={\eps@scaling\columnwidth}]{#2}
 \includegraphics[width={\eps@scaling\columnwidth}]{#3}
}}
\newcommand\plotrtwo[2]{{
 \typeout{Plotrtwo included the files #1 #2}
 \centering
 \leavevmode
 \columnwidth=.5\columnwidth
 \includegraphics[angle=90, width={\eps@scaling\columnwidth}]{#1}
 \includegraphics[angle=90, width={\eps@scaling\columnwidth}]{#2}
}}

\shorttitle{N5813 FROM CHANDRA}
\shortauthors{RANDALL ET AL.}
\slugcomment{}

\begin{document}

\title{Shocks and Cavities from Multiple Outbursts in the Galaxy Group
  NGC~5813: A Window to AGN Feedback}

\author{S.\ W.\ Randall\altaffilmark{1}, W. R. Forman\altaffilmark{1},
  S. Giacintucci\altaffilmark{1,2}, P. E. J. Nulsen\altaffilmark{1},
  M. Sun\altaffilmark{3},
  C. Jones\altaffilmark{1},
  E. Churazov\altaffilmark{4,5}, L. P. David\altaffilmark{1},
  R. Kraft\altaffilmark{1}, M. Donahue\altaffilmark{6},
  E. L. Blanton\altaffilmark{7}, A. Simionescu\altaffilmark{8},
  N. Werner\altaffilmark{8}}

\altaffiltext{1}{Harvard-Smithsonian Center for Astrophysics, 60
  Garden St., Cambridge, MA 02138, USA; srandall@cfa.harvard.edu}
\altaffiltext{2}{INAF/IRA, via Gobetti 101, I-40129 Bologna, Italy }
\altaffiltext{3}{Department of Astronomy, University of Virginia, P.O. Box 400325, Charlottesville, VA 22901, USA}
\altaffiltext{4}{Max-Planck-Institut f\"{u}r Astrophysik,
  Karl-Schwarzschild-Strasse 1, 85741 Garching, Germany}
\altaffiltext{5}{Space Research Institute (IKI), Profsoyuznaya 84/32,
  Moscow 117810, Russia}
\altaffiltext{6}{Physics \& Astronomy Department, Michigan State
  University, East Lansing, MI 48824-2320, USA}
\altaffiltext{7}{Institute for Astrophysical Research and Astronomy
  Department, Boston University, 725 Commonwealth Avenue, Boston, MA
  02215, USA} 
\altaffiltext{8}{KIPAC, Stanford University, 452 Lomita Mall,
  Stanford, CA 94305, USA}

\begin{abstract}

We present results from new {\it Chandra}, {\it GMRT}, and {\it SOAR}
observations of NGC~5813, the dominant central galaxy in a nearby
galaxy group.  The system shows three pairs of collinear
cavities at 1~kpc, 8~kpc, and 20~kpc from the central source, from three
distinct outbursts of the central AGN, which occurred $3 \times
10^6$, $2 \times 10^7$, and $9 \times 10^7$~yr ago.
The H$\alpha$ and X-ray observations reveal filaments of cool gas
that has been uplifted by the X-ray cavities.
The inner two
cavity pairs are filled with radio emitting plasma, and each pair is
associated with
an elliptical surface brightness edge, which we unambiguously identify
as shocks (with measured temperature jumps) with Mach numbers of $M
\approx 1.7$ and $M \approx 1.5$ for the inner and outer shocks, respectively.
Such clear signatures from three distinct AGN outbursts in an
otherwise dynamically relaxed system provide a unique opportunity to study
AGN feedback and outburst history.   The mean power of the two most
recent outbursts differs by a factor of six, from 1.5--10$\times
10^{42}$~erg~s$^{-1}$, indicating 
that the mean jet power changes significantly over long
($\sim10^7$~yr) timescales.  The total energy output of the most
recent outburst is also more than an order of magnitude less than the total
energy of the previous 
outburst ($1.5 \times 10^{56}$~erg versus $4 \times 10^{57}$~erg ), which may be a result of the lower mean power, or may indicate
that the most recent outburst is ongoing.  
The outburst interval implied by both the shock and cavity ages
($\sim10^7$~yr) 
indicates that, in this system, shock heating alone is sufficient to
balance radiative
cooling close to the 
central AGN, which is the relevant region for regulating feedback
between the ICM and the central SMBH.

\end{abstract}

\section{Introduction} \label{sec:intro}

A major result from early {\it Chandra} and {\it XMM-Newton}
observations was that the amount of gas in cool core clusters
that cools to low temperatures is less than predicted by
classical radiative cooling models (David \etal\ 2001; Peterson \etal\
2001; Peterson \& Fabian 2006). 
The implication is that the central gas must be re-heated.  The
source of this heating, and understanding when and how it takes place,
has recently been a major topic of study. A promising candidate is
feedback from energy injection 
by the central AGN of the cD galaxy (\eg, Churazov \etal\ 2001;
Churazov \etal\ 2002; for a review see McNamara \&
Nulsen 2007).
However, the details of this interaction, and how the energy is
transferred from the jets to the ambient ICM, are poorly understood.
Galaxy groups provide an excellent opportunity to study heating and
other non-gravitational processes in the ICM.  Although not as X-ray
luminous as clusters, the effects of heating are more readily apparent in
groups, due to their lower gas temperatures, masses, and central densities.  For
example, the 
gas to total mass fraction in groups ranges from 0.02--0.07, with a
scatter of $\sim2$ 
at a fixed temperature within $r_{2500}$. The scatter is
tightly correlated with the central entropy (Gastaldello \etal\ 2007;
Sun \etal\ 2009),
reflecting the greater role of non-gravitational 
processes in the centers of groups as compared to clusters.

In this paper, we report on {\it Chandra} observations of NGC~5813 (UGC~09655), a
bright ($M_V = -22.01$, Lauer \etal\ 2007) E1 galaxy.  It is the
central dominant member
of a subgroup
(which we shall call the NGC~5813 galaxy group) in the NGC~5846 galaxy
group (Mahdavi \etal\ 2005).  
NGC~5846 and NGC~5813 have a projected separation of 79.7\arcmin\ (740~kpc).
This group is relatively isolated, lying
well off the plane of the Local Supercluster.
Both the NGC~5813 and NGC~5846 galaxy groups are members of the 
  {\it ROSAT}-ESO Flux-Limited X-ray (REFLEX)
galaxy cluster catalog (B\"{o}hringer \etal\ 2004).
Detailed {\it Hubble Space Telescope} ({\it HST}) observations reveal
  that NGC~5813 contains a relatively
undisturbed dusty circumnuclear disk (Tran \etal\ 2001), suggesting
that this galaxy has not recently experienced a major merger.
Emsellem \etal\ (2007) classify NGC~5813 as a ``slow rotator'' galaxy
which, they argue, represents the extreme evolutionary end point reached
in deep potential wells, consistent with this object being a
dynamically old galaxy at the center of a galaxy subgroup.
NGC~5813 contains a $2.8 \times 10^8 M_\odot$ supermassive black hole (SMBH),
and an associated AGN that is a known radio source (\eg, Balmaverde \& Capetti
2006; Magliocchetti \& Br\"{u}ggen 2007).

We report here on a 150~ksec combined observation of NGC~5813 with the
{\it Chandra X-ray Observatory}, an analysis of archival
multi-frequency {\it Very Large Array} ({\it VLA}) observations, 
H$\alpha$ observations with the {\it Southern Astrophysics Research
  Telescope} ({\it SOAR}),
and on some initial results from low
frequency {\it Giant Metrewave Radio Telescope} ({\it GMRT}) radio
observations.  We focus on three main results:

\begin{enumerate}

\item
The ICM in NGC~5813 shows clear signatures from three distinct AGN
outbursts, with three pairs of roughly collinear cavities and
unambiguous shocks with measured
temperature jumps associated with the 
inner and intermediate cavities. 

\item
The mean power of the two most recent outbursts differs by a factor of
six, from 1.5--10$\times 10^{42}$~erg~s$^{-1}$,
even in this otherwise dynamically relaxed system, indicating that the
mean jet power varies over long ($\sim10^7$~yr) timescales.

\item
The heating from shocks alone is sufficient to offset radiative
cooling locally (within at least 10~kpc), without requiring the
internal energy of the X-ray
cavities.  The heating is roughly isotropic, and strongest near the
AGN where the shock Mach 
numbers are larger, which is the region of interest for regulating
feedback between the ICM and the central SMBH.

\end{enumerate}

We assume an angular diameter distance to NGC~5813 of 32.2~Mpc (Tonry
\etal\ 2001), which gives a scale of 0.15~kpc/\arcsec\ for $\Omega_0 = 0.3$,  
$\Omega_{\Lambda} = 0.7$, and $H_0 = 70$~\kms~Mpc$^{-1}$.  All
uncertainty ranges are 68\% confidence intervals (\ie, 1$\sigma$), unless
otherwise stated.

\section{Observations and Data Reduction} \label{sec:obs}

\subsection{{\it Chandra} Observations}\label{sec:chandra}

NGC~5813 was originally observed with {\it Chandra} on April 2,
2005, for 49~ksec (ObsID 5907) with the {\it Chandra} CCD Imaging Spectrometer
(ACIS), pointed such that the galaxy was centered on the
back-side illuminated ACIS-S3 CCD (the ACIS-S1 as well as the
front-side illuminated ACIS-I3 and ACIS-S2 CCDs were also active).  It
was subsequently observed for 
100~ksec on June 5, 2008 (ObsID 9517) with the same chip configuration. These
data were reduced using the method described in
Randall \etal\ (2008).  All data were reprocessed from the
level 1 event files using the latest calibration files (as of {\sc
  CIAO4.2}). CTI and time-dependent gain corrections were
applied. {\sc lc\_clean} was used to remove background
flares\footnote{\url{http://asc.harvard.edu/contrib/maxim/acisbg/}}.
The mean event rate was calculated from a source free region using
time bins within 3$\sigma$ of the
overall mean, and bins outside a factor of 1.2 of this mean were
discarded.  There were no periods of strong background
flares.  The resulting cleaned exposure times were 48.7 and 99.6~ksec,
respectively.

Diffuse emission from NGC~5813 fills the image FOV for each
observation.  We therefore used the {\sc
  CALDB\footnote{\url{http://cxc.harvard.edu/caldb/}}} 
blank sky background files appropriate for each observation (including
the new ``period E'' files for the more recent observation),
normalized to our observations in the 10-12 keV energy band.
To generate exposure maps, we used a MEKAL model with $kT = 0.7$~keV,
Galactic absorption, and abundance of 30\% solar at a redshift $z =
0.006578$, which is consistent with typical results for the extended
emission from detailed
spectral fits (see \S~\ref{sec:spec}).

The exposure corrected, background subtracted, 0.3--2~keV {\it
  Chandra} image is shown in Figure~\ref{fig:fullimg}. To enhance the
  visibility of the diffuse emission, bright point sources were
  removed, and the regions containing point sources were ``filled in''
  using a Poisson distribution whose mean was equal to that of a local
  annular background region.  A close-up of the core in
  Figure~\ref{fig:xcore} shows small-scale structure
  in the center, while the more heavily smoothed image in
  Figure~\ref{fig:oedge} shows structure in the fainter outer
  regions. We discuss the main features in these images in
  \S~\ref{sec:ximg}.

All X-ray spectra were fitted in the 0.6--3.0~keV band and grouped to a
minimum of 40 counts per spectral bin.  The absorption was fixed to
the Galactic value of $N_H = 4.37\times 10^{20}$~cm$^{-2}$ (Kalberla
\etal\ 2005).  Varying the absorption from the Galactic value did not
significantly improve the fit.  Anders \& Grevesse (1989) abundance
ratios were assumed throughout, unless otherwise stated.
Temperature maps were derived using the method of Randall \etal\
(2008).  For each 
temperature map pixel, we extracted a spectrum from a circular region
containing a minimum number of net counts (after subtracting the blank sky
background) in the 0.6 -- 3.0~keV band. The resulting spectrum
was fitted with an absorbed {\sc apec} model using
{\sc xspec}, with the abundance allowed to vary.

\subsection{Radio Observations}\label{sec:radio}

NGC~5813 was observed with the {\it GMRT} at 235 MHz, as part of a larger, 
ongoing project (Giacintucci \etal\ 2009;
Giacintucci \etal\ 2010). The observations were carried out in August 2008
for a total of 100 minutes on source. The NRAO Astronomical Image
Processing System (AIPS) package was used for the data reduction and
analysis. The visibilities were 
inspected and edited to identify and remove bad data. After the initial
calibration, phase-only self-calibration was applied to remove residual
phase variations.
Multi-field imaging was implemented in each step of the data
reduction. The rms noise level (1$\sigma$) achieved in the image
is ~0.3 mJy/beam. We refer to Giacintucci \etal\ (2010)
for a detailed description of the data reduction and analysis.

We also include results from an analysis of archival {\it VLA}
observations at multiple frequencies. Data
calibration and imaging were carried out in AIPS following the standard
procedure (Fourier-transform, Clean and Restore). Phase-only
self-calibration was applied to remove residual phase variations and
improve the quality of the image.
The properties of the {\it VLA} and {\it GMRT} radio observations are
summarized in Table~\ref{tab:radio}.

\subsection{H$\alpha$ Observations}\label{sec:halpha}

Narrow-band H$\alpha$ imaging observations were made with
the {\it SOAR} Optical Imager (SOI) on July 6, 2008 (UT).  The night
was clear and photometric.
Two CTIO narrow-band filters were used, 660075-4
for the H$\alpha$+[NII] lines and 6129/140 for the continuum.
Three 15-minute exposures were taken with the 660075-4
filter and three 12-minute exposures were taken with the
6129/140 filter. Spectroscopic standard
stars were EG274 and LTT7987. More detail on the SOI
data reduction can be found in Sun et al. (2007).

\section{X-ray Cavities and Shocks} \label{sec:ximg}

As we discuss in detail in this paper, the X-ray images show signatures from
 three episodes of AGN activity.  There are two clear pairs of
 cavities distributed collinearly (in a direction roughly parallel to the minor
  axis of the elliptical optical isophotes) and symmetrically about
 the galaxy center, 
 each pair associated with a shock.  In
 addition, we argue for a third pair of cavities at larger radii
 (along roughly the same line) and an associated third surface
 brightness edge, which may be a weak shock.
 In detail, we see the following:

\begin{enumerate}

\item
A pair of inner cavities at $\sim 1.4$~kpc, inflated by the most recent AGN
outburst about $3 \times 10^6$~yr ago (Figure~\ref{fig:xcore}).
The cavities are surrounded by bright rims of emission, similar to
what is seen in other systems (\eg, the Perseus Cluster 
Fabian \etal\ 2003, M87 Forman \etal\ 2007; M84 Finoguenov \etal\
2008, Abell~2052 Blanton 
\etal\ 2009; see Figure~\ref{fig:xcore}), though here we find the rims
to be hotter than the ambient gas, whereas in other systems they are
typically cooler.
We identify the sharp edge in the rims, 1.4~kpc southeast of
the AGN, as a shock, which we refer to as the 1.5~kpc shock, with a
Mach number of $M = 1.7$ (see \S~\ref{sec:shocks}).
The rims overlap to
form an indented structure to the northwest.  The northeastern inner cavity has
an irregular morphology, possibly due to the wall of the cavity being
``punched through'' by the AGN jet.

\item
A pair of intermediate cavities at $\sim 8$~kpc, inflated by a previous AGN
outburst about $2 \times 10^7$~yr ago (Figure~\ref{fig:fullimg},
Figure~\ref{fig:cavoverlay}). The southwestern cavity has a regular
morphology, while 
the northeastern cavity is more extended in the radial direction and
may be a double ``Russian doll'' cavity (\eg, as in M84, Finoguenov
\etal\ 2008). The intermediate cavities
lie just inside a sharp, elliptical edge, which we identify as a
shock (and refer to as the 10~kpc shock) with  $M = 1.5$ (see
\S~\ref{sec:shocks}).

\item
A faint pair of outer cavities, at $\sim 20$~kpc, inflated by a
previous outburst about $9 \times 10^7$~yr ago
(Figure~\ref{fig:oedge}, Figure~\ref{fig:cavoverlay}).  The
northeastern outer cavity is surrounded 
by a rim of brighter emission, while the southwestern outer cavity is
a weaker feature (see \S~\ref{sec:cavities}).  The outer cavities are
associated with a weak outer 
edge-like feature at $\sim 25$~kpc (Figure~\ref{fig:oedge}), just past
the outer edges of the cavities.  This feature may represent an old
shock, or a transition region from the galaxy atmosphere to the
extended group atmosphere (see \S~\ref{sec:outer}).

\item
A central point source, offset $\sim 0.5$~kpc southeast of the axis
defined by the 
cavities (Figure~\ref{fig:xcore}), with an X-ray luminosity of $L_X =
1.6 \times 10^{39}$~erg~s$^{-1}$ in the 0.3 -- 12.0~keV rest frame
energy band.  Although this source is quite faint, we identify it as the AGN that has inflated the
X-ray cavities and driven the shocks in the ICM due to its central location and
detection in the radio (see \S~\ref{sec:agn})

\end{enumerate}
Since it is difficult to show all the cavities simultaneously in the
same image, we divided the X-ray image by a $\beta$-model to better
show surface brightness fluctuations over a wider dynamic
range.  The resulting image is shown in Figure~\ref{fig:cavoverlay},
with the cavities indicated by overlaid regions for clarity.
The overall morphology of NGC~5813 is remarkably symmetric and
  regular, suggesting that  AGN feedback maintains a near ``steady
  state'' through regular outbursts in an otherwise
  undisturbed system (consistent with optical studies that also
  conclude NGC~5813 is dynamically old, Tran \etal\ 2001; Emsellem \etal\ 2007).

\subsection{Radio Emission from the X-ray Cavities} \label{sec:rimg}

Radio contours from our 235~MHz {\it GMRT} observations (green) and archival
1.36~GHz {\it VLA} observations (blue) are 
shown overlaid on the {\it Chandra} image of the core in
Figure~\ref{fig:core_img}.  
Extended radio emission at 1.36~GHz fills the inner cavities, while
the 235~MHz emission overfills the cavities, extending along
the axis of symmetry of the cavities (this extension is real, and not
due to the larger beam size at 235~MHz).
The intermediate cavities are filled with 235~MHz radio
emission, but show no emission at 1.36~GHz.  
The intermediate cavities are also not detected at 1.4~GHz in the {\it
  NRAO VLA Sky Survey} ({\it NVSS}, which has a beam size of 45\arcsec
) or at 1.36~GHz in archival {\it
  VLA} C array configuration observations (see Table~\ref{tab:radio}).
The outer cavities, which
are outside the FOV of Figure~\ref{fig:core_img}, do not show any
detected radio emission.
The radio images are therefore qualitatively consistent with what is
expected for intermittent 
AGN outbursts, where the electrons contained in the cavities age
due to synchrotron, inverse Compton, and adiabatic losses as the 
cavities rise buoyantly after the outburst phase.  The central
cavities contain recently accelerated 
electrons, which emit at high and low frequencies, while older
cavities contain older electron populations with fewer energetic
particles and weaker high frequency emission.

\section{The Thermal Structure of the Gas} \label{sec:spec}

\subsection{Temperature Map} \label{sec:tmap}

\begin{figure}
\plotone{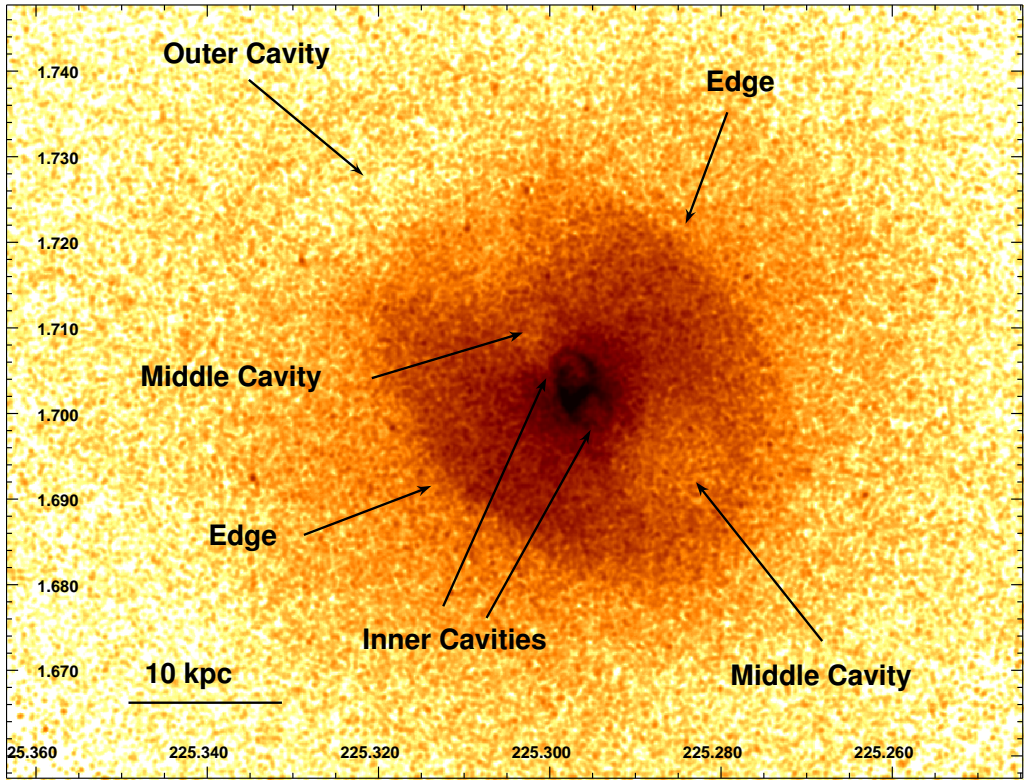}
\caption{
  Exposure corrected, background subtracted 0.3--2
  keV {\it Chandra} image of NGC~5813.  The image has
  been smoothed with a 1.5\arcsec\ radius Gaussian and point sources
  have been filled-in by randomly drawing from a Poisson distribution
  fit to a local background annular region.
  The image shows two pairs of cavities, plus an outer cavity to the
  northeast, two sharp edges to the northwest and southeast, and
  bright rims around the pair of inner cavities.
  \label{fig:fullimg}
}
\end{figure}

\begin{figure}
\plotone{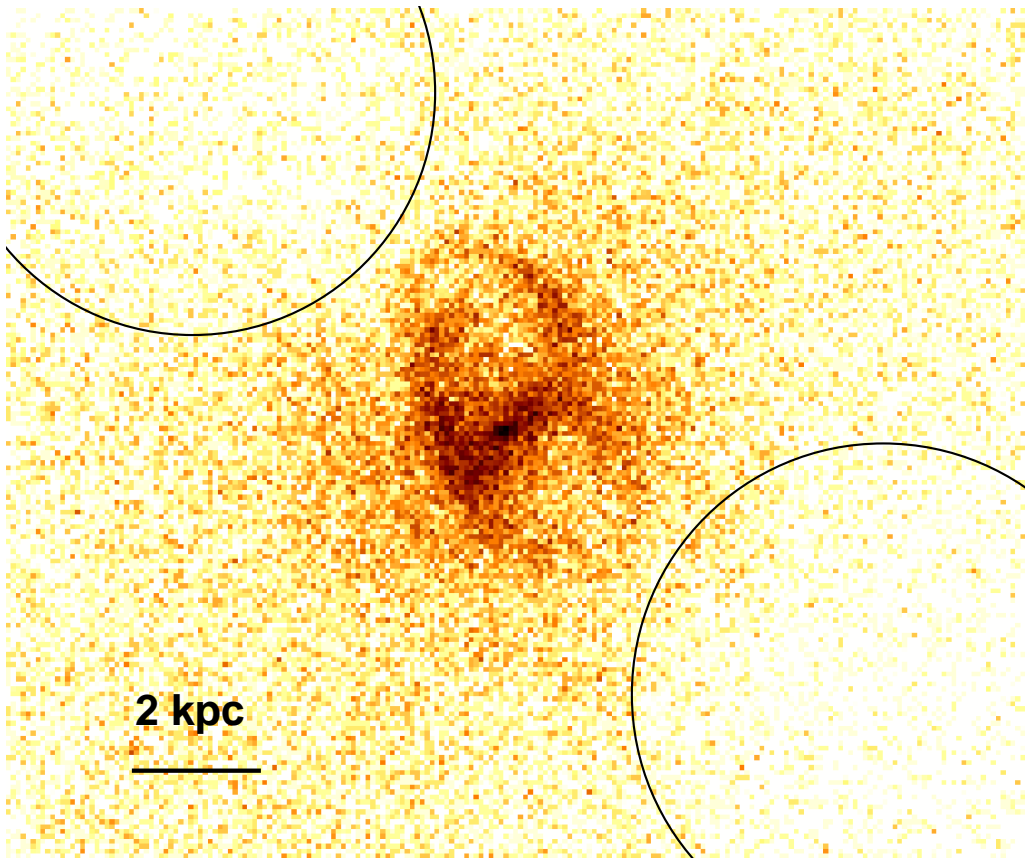}
\caption{The core from the image in
  Figure~\ref{fig:fullimg}, unsmoothed.
  The point-like
  AGN is visible, offset $\sim 0.5$~kpc southeast of the axis of
  symmetry of the inner cavities.  There is a sharp edge just southeast of
  the AGN, and the two rings surrounding the inner cavities overlap to
  form an indented structure to the northwest.  The black circles define
 the inner edges of the intermediate cavities.
  \label{fig:xcore}
}
\end{figure}

The X-ray image (Figure~\ref{fig:fullimg}) shows complicated structure
in the ICM, which fills the FOV.  To study the thermal
structure of the ICM, we generated a temperature map, requiring 1500
net counts per extraction region.
 The resulting temperature map, with the X-ray cavity regions
 overlaid, is shown in Figure~\ref{fig:tmap}.
The corresponding pseudo-pressure and pseudo-entropy maps are shown in
Figure~\ref{fig:press}.
The extraction radii range from 2.8\arcsec\ (0.4~kpc) in bright
regions near the core to 59\arcsec\ (9~kpc) in faint outer regions.
The temperature uncertainties are between 2\% -- 3\% across the map.

The temperature map shows that even in the
projected, effectively smoothed map, the hot (0.7--0.75~keV) 10~kpc
shocks are visible 
at the location of the prominent surface brightness edges.  
The pseudo-pressure map also shows large jumps across the edges,
consistent with these features being shock fronts.
There is a
trail of cool 0.55~keV gas though the galaxy center, along the line
defined by the X-ray cavities
indicated in Figure~\ref{fig:fullimg}, terminating at the edges of
the intermediate cavities (we discuss this feature further in
\S~\ref{sec:uplift}). 
The  $kT \sim 0.65$~keV gas extends to larger radii (out to $\sim
27$~kpc) in the
east-northeast, coincident 
with the extension of diffuse emission across the outer edge in
Figure~\ref{fig:oedge}.  
East of this extension, the temperature
rises rapidly from about 0.65~keV to 0.75~keV over $\sim 7$~kpc.

To study the detailed structure in the core, we made a
higher angular resolution temperature map of the central region of NGC~5813.
In addition to the finer spatial
binning, each extracted spectrum had only 1000 net counts, giving
smaller extraction radii (and thus less smoothing in the map), ranging
from 0.4~kpc to 1.8~kpc. 
The resulting temperature map is shown in Figure~\ref{fig:tmap_core},
and the corresponding pressure and entropy maps in
Figure~\ref{fig:core_press}, 
with the 0.3--2.0~keV X-ray surface brightness contours overlaid.
The bright rims surrounding the innermost bubbles are revealed to
contain relatively hot ($kT \approx 0.7$~keV), high pressure gas, in
contrast to the cool
bubble rims seen
in some other systems (\eg,  the Perseus
Cluster, Fabian \etal\ 2003; M87, Forman \etal\ 2007; Abell~2052,
Blanton \etal\ 2009).  The gas in the 
rims has likely been shock heated by a recent AGN
outburst, which has rapidly inflated the innermost cavities.

\subsection{Detailed Spectral Analysis} \label{sec:dspec}

\subsubsection{Azimuthally Averaged Profiles} \label{sec:profs}

We produced projected radial profiles by fitting spectra extracted
from concentric annuli, centered on the centroid of the diffuse emission
at larger radii.  Each annular bin was fitted with an absorbed {\sc apec}
model, with the abundance allowed to vary.  The resulting temperature
profile is shown in the top panel of Figure~\ref{fig:azprof}.  There is a 
temperature increase of $\sim 0.05$~keV at $\sim$10~kpc, at the
location of the surface brightness edges
indicated in Figure~\ref{fig:fullimg}, unambiguously identifying these
features as shocks.  Additionally, there is an inner temperature jump
of $\sim 0.06$~keV
at $\sim$2~kpc, at the location of the bright rims around the
inner bubbles, consistent with the core temperature map shown in
Figure~\ref{fig:tmap_core}.  Although the point source presumed to be
the central AGN has been excluded, the temperature profile shows a
strong central spike.  
This is because the profile center is in the region of the hot
overlapping rims of the central cavities, shown in Figure~\ref{fig:xcore}.

\subsubsection{Deprojection Analysis} \label{sec:deproj}

To determine the 3D structure of the ICM we performed a deprojection
analysis using concentric annuli as in \S~\ref{sec:profs}, but with
bins 2--3 times larger to provide adequate statistics for the
deprojection analysis.  We used the
``onion peeling'' method (employed, \eg, by Fabian \etal\ 1981,
Blanton \etal\ 2003) to 
derive deprojected profiles.
First, the projected temperature,
abundance, and {\sc xspec} normalization are determined by fitting an
absorbed {\sc apec} model to the outermost annulus.  
Fits to spectra from annuli at smaller radii are then determined by
adding an additional component for each outer annulus, with fixed
temperature and abundance, and a normalization scaled to project from
the outer to the inner annulus, assuming spherical symmetry.  
We note that this procedure does not correctly account for the
uncertainties, since 
the contributions from outer shells are fixed.
This method was adopted instead of simultaneously fitting spectra in
all of the bins (\eg, with the {\sc projct} model in {\sc xspec}) to
prevent spectra at small radii, where the
spherical symmetry approximation is less accurate, influencing the
fitting results at large radii.
A comparison of the two methods shows that while the best-fitting
temperature profile is not significantly affected, 
the density profile determined with the {\sc projct} model differs
somewhat from the profile we present here and shows a
$\sim$50\% increase in
density with radius between 6--9~kpc (in the region of the
intermediate bubbles and 10~kpc shock front).  Thus, while the azimuthally
averaged deprojected temperature
profile is not significantly affected by the assumption of spherical
symmetry, the density profile is sensitive to this assumption and is
therefore only approximate.

As a further check on the deprojection, we performed an independent
analysis, with the data analyzed as described in Vikhlinin \etal\
(2005) and the deprojection method described in Churazov \etal\ (2008,
2010).  This method accounts for emission at large radii by assuming a
power law density profile for emission outside of the FOV and fitting
the normalization of this
component along with the fluxes from spherical shells.  The derived
temperatures agreed within the 1$\sigma$ confidence range, and the
temperature jumps were fully consistent within the uncertainties.

In the case of NGC~5813, measuring the deprojected abundance
profile is extremely difficult.  This is because at these relatively
low gas temperatures (0.6--0.7~keV) the spectral resolution of the
ACIS instrument is such that there is a
strong degeneracy between line emission (which determines the
abundance) and continuum emission (which determines the electron
density).  When this effect is accumulated across several shells
and projected onto the inner shells, the inner abundances are essentially
indeterminate, leading to a highly uncertain density profile. 
Furthermore, determining abundances in multiphase gas is a known
problem (\eg, Buote 1999; Rasia \etal\ 2008).
We therefore fixed the abundance at 50\% solar, which is an average
value from the projected profile fits.

The resulting deprojected (red triangles) and projected (black
circles) temperature profiles are shown in
the top panel of Figure~\ref{fig:azprof}, alongside the corresponding
density, pressure, and entropy profiles.  
The entropy is taken to be $S = kT n_{\rm e}^{-2/3}$ and the pressure
$P = nkT$, where $n_{\rm e}$ is the 
electron density and $n = 1.8 n_{\rm e}$.
The most significant effect of the deprojection on the temperature
profile is
the lower temperature of the gas at $\sim$5~kpc (0.55~keV versus
0.61~keV) once the projection
effects from the outer shock-heated gas have been removed, as well as a
significant increase in the central temperature (from 0.65~keV to
0.7~keV).  Note that in these plots
the shock front edges are smeared over multiple bins
since they are not spherically symmetric with respect to the
galaxy center.

\subsubsection{Temperature Profile Across the Shocks} \label{sec:edges}

The azimuthally averaged temperature profiles
(Figure~\ref{fig:azprof}) and projected temperature (Figure~\ref{fig:tmap})
map show temperature rises at the location of the surface brightness
edges, characteristic of shock fronts.  We therefore extracted spectra
in sectors across these edges to better characterize the temperature
jumps.  Each sector was centered on the center of curvature defined by
the corresponding edge, which is {\it not} coincident with the central
position used to extract the azimuthally averaged profiles.  
The extraction regions were truncated at small radii to avoid the
complex structure in the central region (additionally, the assumption
of spherical symmetry breaks down at small radii since the centers of
curvature are not coincident with the overall centroid of the diffuse
emission).
The projected and
deprojected temperature profiles are shown in
Figure~\ref{fig:sec_ktprof}.  
The northwestern region shows higher overall temperatures and a larger
temperature jump of $\sim 0.15$~keV as compared to $\sim 0.1$~keV in
the southeast 
(where we estimate the size of the jumps by extrapolating the
base-line deprojected temperature increase on either side of the
temperature peak, roughly from $r<10$~kpc and $r>18$~kpc).
The lower temperatures in the southeast are likely due in part to the
east-northeast extension of cool gas seen in the temperature map
(\S~\ref{sec:tmap}), which overlaps with the extraction annuli.

\subsubsection{Total Diffuse Emission} \label{sec:total}

To accurately determine the weighted average properties of
the diffuse gas, we extracted and fitted spectra for the total diffuse
emission within 2.9\arcmin\ (27~kpc).  We initially fitted the spectra
with a single {\sc apec} model.  The resulting model showed residuals
near the 1.8~keV Si and 2.46~keV S lines, possibly indicating
non-solar abundance ratios.  We therefore fit the spectra with a {\sc
  vapec} model with the abundances of O, Ne, Mg, Si, S, and Fe allowed
to vary independently (other elements were not well constrained and
were fixed at 1/2 the solar abundance, except He which was fixed at
solar).  
The resulting fitted abundance values were 
0.8 -- 1.0 solar, with Si and S having somewhat larger values than Ne
and Mg, except for O ($Z_O = 0.35 \pm 0.03$) and Fe
($Z_{Fe} = 0.65 \pm 0.03$).  Since the total diffuse emission is
expected to include emission from gas at multiple temperatures, we
added a second {\sc vapec} component to the model and tied individual
abundances for the two components together.  This model provided a
much improved fit, 
with abundance values of $Z_O = 0.13^{+0.03}_{-0.02}$,  
$Z_{Ne} = 0.59^{+0.05}_{-0.04}$, $Z_{Mg} = 0.65 \pm 0.04$,
$Z_{Si} = 0.78 \pm 0.04$, $Z_S = 0.94 \pm 0.08$, and $Z_{Fe} = 0.53 \pm 0.03$.
The fitted temperatures were $kT_1 =
0.35^{+0.03}_{-0.01}$ and $kT_2 = 0.671^{+0.011}_{-0.007}$.  
We tried including a power-law component in each of the above models, but in
no case did this addition improve the fit (even if the photon index
was fixed at a typical value of 1.5 for an unresolved LMXBs
population) or tightly constrain the 
photon index.  This suggests that emission from unresolved LMXBs is not
significant in this energy band.
We note that the above results depend somewhat on the
adopted abundance table.  If we adopt the updated abundance table of Grevesse
\& Sauval (1998), we find that Fe is no longer under-abundant
compared to the other elements, with $Z_{Fe}$, $Z_S$, and $Z_{Si}$ all
roughly 0.7 solar, and  $Z_{Ne}$ and  $Z_{Mg}$ about 0.6 solar.  O is
still found to be under-abundant, with $Z_O = 0.16^{+0.04}_{-0.03}$.
This difference in the fitted value for $Z_{Fe}$ was noted previously,
\eg, by Humphrey \etal\ (2004).

As a further check of this result, we examined data from {\it XMM-Newton}
RGS observations, which has better spectral resolution in the region
of the O lines.
The processing of the RGS data is
described in Werner \etal\ (2009).  We fitted the spectrum from a 1\arcmin\ wide region centered on the core of
NGC~5813 between 10--25~\AA\ with a
collisionally ionized plasma ({\sc cie}) model with two cooling flow
components, one to account for gas cooling down to 0.4~keV and a
second to account for gas cooling to lower temperatures.  The
abundances were constrained to be equal across each model component.
We found best-fitting abundance values of $Z_{Fe} = 0.60 \pm 0.06$,
$Z_O = 0.44 \pm 0.06$, and $Z_{Ne} = 0.42 \pm 0.06$.  Although this
gives a higher $Z_O/Z_{Fe}$ ratio (0.7 versus 0.2 from {\it Chandra}),
O is still found to be
under-abundant as compared to Fe.
We conclude that there is evidence for
non-solar abundance ratios in the diffuse ICM, with the most robust
result being a decreased O abundance relative to solar.
Similar sub-solar values for $Z_O/Z_{Fe}$ have been reported for other
galaxy groups, massive elliptical galaxies, and clusters of galaxies,
and imply a greater relative enrichment from Type Ia supernovae as
compared to Type II (\eg, Finoguenov \etal\ 2000; Finoguenov \etal\ 2001).
However,  sub-solar  $Z_O/Z_{Fe}$ values are difficult to reconcile
with the larger $Z_{Si}/Z_{Fe}$ and 
$Z_S/Z_{Fe}$ abundance ratios when one tries to apply SN enrichment models to
explain the observed ICM abundances
(see Humphrey \etal\ 2004 and references therein for a discussion).

\subsubsection{The Central Source} \label{sec:agn}
NGC~5813 contains a central X-ray point source, visible in
Figure~\ref{fig:xcore}.
We extracted a spectrum for the central source using an
aperture with a radius of 1.5\arcsec and a background determined from
a local annular region. The spectrum, which had 550 counts in the
0.6--5.0~keV band, was fitted with an absorbed
power-law, giving a best-fitting photon index of $\Gamma =
1.5^{+0.5}_{-0.6}$.  The nuclear X-ray luminosity is $L_X =
1.6^{+0.8}_{-0.5} \times 10^{39}$~erg~s$^{-1}$ in the 0.3 -- 12.0~keV
rest frame
energy band (90\% confidence ranges).
We calculated the radio luminosity between 10~MHz and 100~GHz using
archival {\it VLA} A array configuration observations of the core at
1.49~GHz and 4.86~GHz.  This gave a spectral index of $\alpha_r =
0.35$ and a luminosity of $L_R =  1 \times 10^{36}$~erg~s$^{-1}$.
Even though the spectral index indicates a relatively flat spectrum
for a core AGN, giving a larger estimate of the broad band luminosity
than would a steeper spectrum, the calculated luminosity is still
two orders of magnitude fainter than the faintest source in the sample
of systems containing X-ray cavities given in B\^{i}rzan \etal\
(2008). However, they mainly consider sources at the cores of galaxy
clusters, which are 
higher mass systems than the subgroup we consider here.

Although the central source is X-ray faint enough to be classified as
an ultraluminous 
X-ray source (ULX), we identify it as the AGN that has inflated the
X-ray cavities and driven the shocks in the ICM due to its central location and
detection in the radio (see Figure~\ref{fig:core_img}).
Its low luminosity identifies this source as a low luminosity
AGN, which typically have $L_X <  1 \times 10^{42}$~erg~s$^{-1}$
(Ptak 2001).  This suggests that the source is either faint yet
mechanically powerful, in a quiescent
state after having recently had an outburst that inflated the inner
X-ray cavities, or heavily obscured (allowing the absorption to
vary gave a value that was 4 times Galactic, although the fit was not
improved and the absorption was consistent with the Galactic value to
within 1$\sigma$).


%

\section{Discussion} \label{sec:discuss}

\subsection{Structure of the Shock Fronts} \label{sec:shocks}

The hard band image, temperature and pressure maps, and deprojected
temperature and pressure profiles identify the prominent surface
brightness edges around the intermediate cavities as shock fronts.  To
quantitatively study
the structure of these shocks, we extracted the 0.3-2.0~keV surface
brightness profiles from the sectors used to derive the temperature
profiles shown in
Figure~\ref{fig:sec_ktprof}.  
Note that the profiles are defined by the centers of curvature 
of the 10~kpc shocks, which do {\it not} coincide with the location of the central AGN.
The profiles were then converted to
integrated emission measure (IEM) profiles, using the temperatures from
the projected temperature profile 
shown in Figure~\ref{fig:sec_ktprof}, with the abundance fixed at 50\%
solar (roughly the average from projected profile fits in this region,
see \S~\ref{sec:deproj}).  The resulting IEM profiles are
shown in Figure~\ref{fig:edges}.  Each profile shows a sharp edge at
$\sim$13~kpc. Following our previous
work (Vikhlinin \etal\ 2001; Randall \etal\ 2009a, 2009b), we fit the
profiles by
projecting a 3 dimensional density profile consisting of two power
laws, connected by a discontinuous break, or ``jump''.  The free
parameters were the normalization, the inner ($\alpha$) and outer
($\beta$) power law slopes, the position of the
density discontinuity ($b_{\rm break}$), and the amplitude of the jump
($A$). The
best-fitting model is shown as the solid lines in
Figure~\ref{fig:edges}, with the fitted break radii indicated by the
vertical dashed lines. 
The best fitting inner density jumps for the northwest and southeast sectors
are $A_{{\rm nw}} = 1.75^{+0.04}_{-0.03}$ and $A_{{\rm se}} =
1.69 \pm 0.07$.  Using the Rankine-Hugoniot shock jump 
conditions for a $\gamma = 5/3$ gas implies Mach numbers of 
$M_{{\rm nw}} = 1.53$ and $M_{{\rm se}} = 1.48$, and temperature
jumps by the same factor as the Mach numbers. 
The
predicted temperature jumps are greater than the temperature
jumps of $\sim 1.2$ detected in the deprojected temperature profiles
(Figure~\ref{fig:sec_ktprof}).  
We discuss this discrepancy below.

The core temperature map (Figure~\ref{fig:tmap_core}) and the
azimuthally averaged temperature profile (Figure~\ref{fig:azprof})
show evidence for shock heated gas in the bright rims surrounding the
innermost bubbles, about $1-2$~kpc from the central AGN.  The X-ray
image of the core (Figure~\ref{fig:xcore}) shows a sharp
contrast between the bright bubble rims and the surrounding gas, with
the sharpest edge to the southeast.  We extracted the surface
brightness profile in a 52\mydeg\ wide
sector across the southeastern edge, centered on the AGN, and fit the
IEM profile with a broken power law density model as above.  To the
northwest, the morphology is more complicated, with the two rims
meeting to form an indented structure.  Since this structure is not
well-modeled by our assumption of spherical symmetry we did not fit
the IEM profile to the northwest.
There were too few counts to accurately measure the
temperature profile across the southeast, so we assumed an isothermal
gas with $kT =
0.6$~keV and 50\% solar abundance.
The resulting IEM profile, along with the best fitting model, is shown
in Figure~\ref{fig:core_emfit}.
The deviant point at 2~kpc is due to small-scale clumpiness in the
ICM, visible in the X-ray image.
We find a density jump of $1.97 \pm 0.12$, which corresponds to
a $M = 1.71 \pm 0.1$ shock with a temperature jump of $1.72 \pm 0.12$.
Unfortunately, 
there are too few counts to derive a deprojected temperature profile
to compare to the inferred temperature jump, although the corresponding
inner jump in the azimuthally averaged temperature profile of about 1.1
(Figure~\ref{fig:azprof}) is much smaller than the jump of 1.7 predicted from
the density discontinuity.  
To better quantify the temperature jump across the 1.5~kpc shock,
we extracted a spectrum of the total emission from the post-shock
region in the core wedge used to fit the surface brightness profile,
and from a similar sized region in the same wedge just outside the
shock edge.  We find that the (inner) post-shock region is best fit by a two
temperature thermal model, with best-fitting temperatures of $kT_1 =
0.64^{+0.03}_{-0.04}$ and $kT_2 = 0.97^{+0.21}_{-0.17}$, while the
pre-shock region is adequately described by a single thermal model
with $kT = 0.5 \pm 0.01$.  If we assume that the two thermal
components in the post-shock region give the temperatures of the shocked
gas and local projected pre-shocked gas this gives a temperature jump
of $1.5 \pm 0.3$.  This is consistent with a temperature jump
of 1.7 as expected from the Mach number, although the large uncertainties limit
the usefulness of this comparison.
The properties of the 10~kpc and 1.5~kpc shocks are summarized in
Table~\ref{tab:shocks}.

For both the 1.5~kpc and 10~kpc shocks, the observed temperature rise is
less than what is expected based on the Mach number, even in the
deprojected temperature profiles.
To better estimate the expected measured temperature rise
associated with the shocks we ran 1D hydrodynamical simulations to fully
model their evolution.  For the 10~kpc shocks, the simulations start
with an isothermal
sphere with $kT = 0.67$ and a power law density profile, in
hydrostatic equilibrium.  The logarithmic slope of the density profile
was taken
to be 1.46, which was 
determined by fitting the slope of the surface brightness profile
beyond the 10~kpc shocks.  The shocks were initiated by a central point
explosion and allowed to propagate freely.  Since the gas is initially
isothermal with a power law density profile, the simulations are scale
free, so we can choose the point of evolution at which the code best
reproduces the observed surface brightness jump at the shock front and
scale accordingly.  The resulting Mach number for the 10~kpc shocks is
$M=1.52$, in excellent agreement with what was found above by fitting a
discontinuous power law density model to the integrated emission
measure profile.  The expected projected temperature profile is shown
in Figure~\ref{fig:hydro_projkt}.  The solid lines show the emission
weighted temperature (with 90\% confidence ranges) and the points show
single temperature fits to more accurately simulated model
spectra folded through the {\it Chandra} response (the discrepancy
between emission weighted temperatures and fitted temperatures for
multi-temperature gas has previously been noted by Mazzotta \etal\ 2004).
The simulations predict a projected temperature rise of
$\sim 0.1$~keV, consistent with observations (see Figure~\ref{fig:azprof}).
Similar simulations fit to the 1.5~kpc shock to the southeast predict a
measured projected temperature rise of $\sim30\%$, consistent with
observations (although the observational uncertainties are large).
We conclude that the projected temperature profiles are
consistent with the calculated Mach numbers (although the temperature
structure of the 1.5~kpc shock is poorly constrained).  The deprojected
temperature measurements cannot resolve the temperature jump due to
the narrow width of the shock and the rarefied cool gas behind
the shock (the temperature of which drops slightly below the ambient
temperature in our isothermal simulations).

\subsection{The Outer Edge at 25~kpc} \label{sec:outer}

\begin{figure}
\plotone{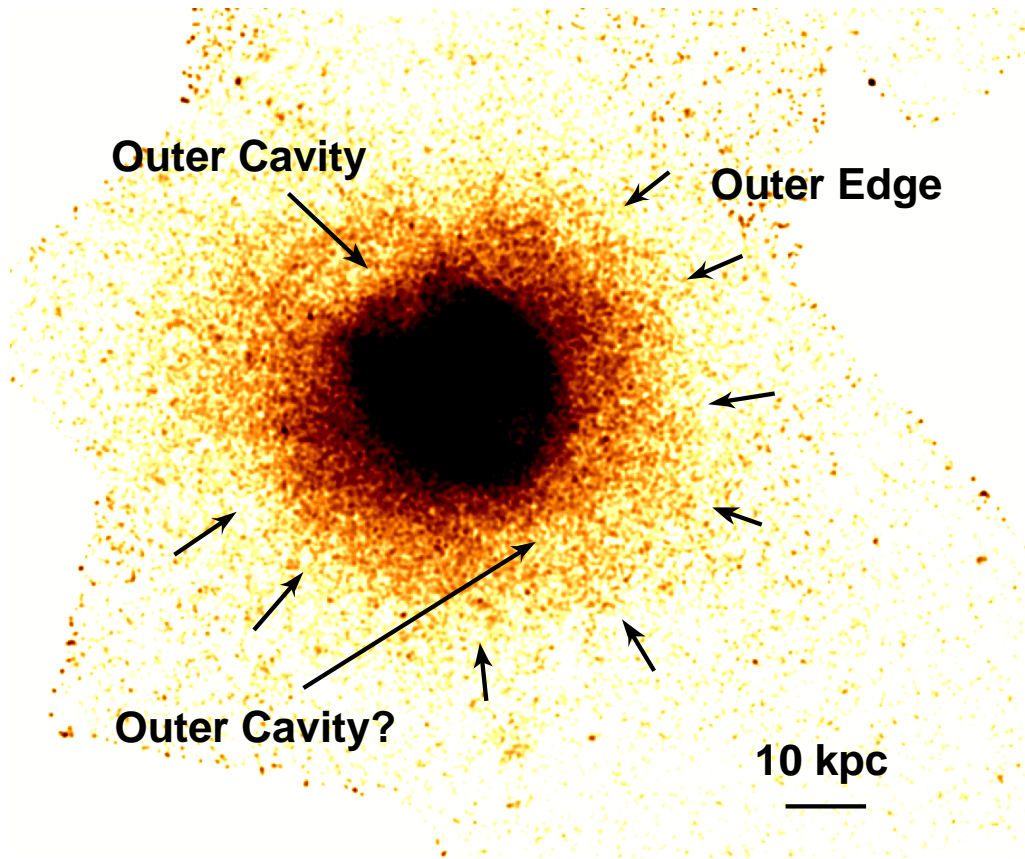}
\caption{
  Same as Figure~\ref{fig:fullimg}, but smoothed with a 3\arcsec\
  radius Gaussian to show faint features in the outer regions.
  The image shows a weak outer cavity to the southwest, and a faint
  outer edge-like feature (indicated by the multiple short arrows)
  that encircles NGC~5813.  To the northeast the edge structure is
  less clear, as there is brighter diffuse emission extending across the
  edge out to larger radii in this direction.
  \label{fig:oedge}
}
\end{figure}

\begin{figure}
\plotone{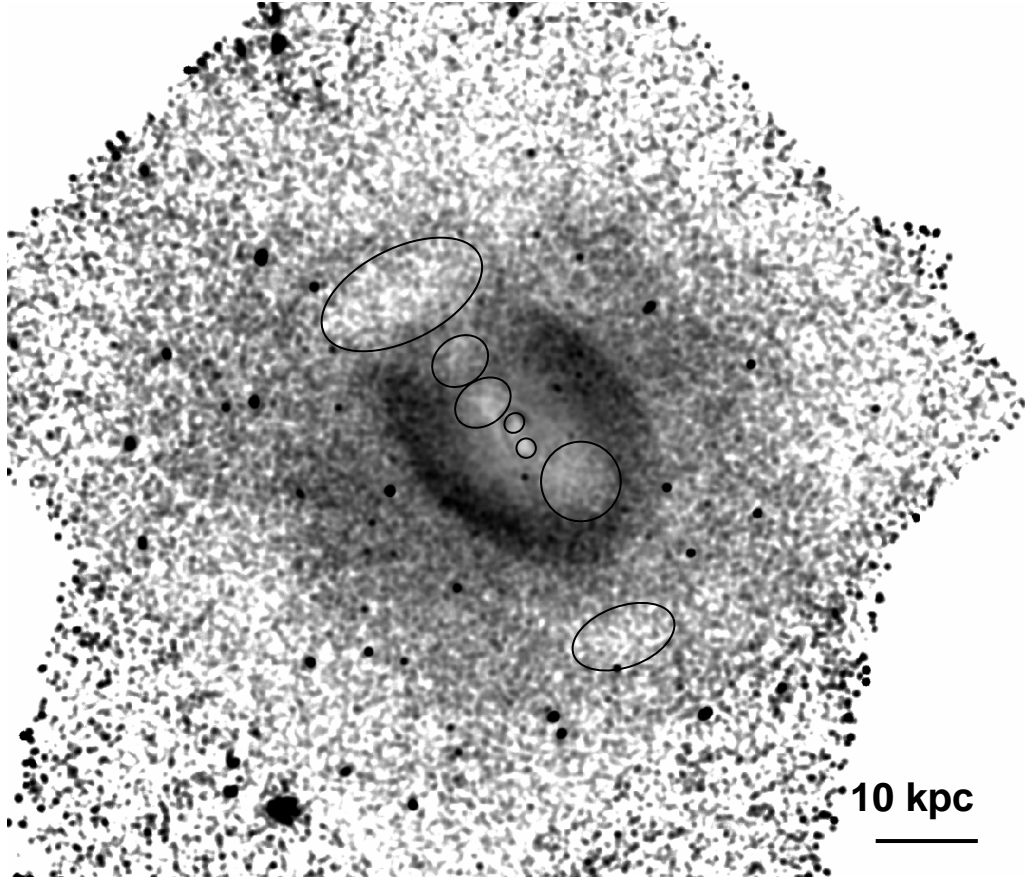}
\caption{
  0.3--2.0~keV {\it Chandra} image from Figure~\ref{fig:fullimg} (with
  sources included),
  divided by a fitted beta model and smoothed with a 6\arcsec\
  Gaussian to more clearly show surface
  brightness fluctuations over a wider dynamic range.
  The cavity regions listed in
  Table~\ref{tab:cavities} are overlaid for clarity.  We see the two
  small inner cavities at $\sim 1.5$~kpc, the middle cavities (with
  the northeastern 
  ``Russian doll'' cavity split into two piece) at $\sim 8$~kpc, and
  the outermost cavities at $\sim 20$~kpc.
  \label{fig:cavoverlay}
}
\end{figure}

\begin{figure}
\plotone{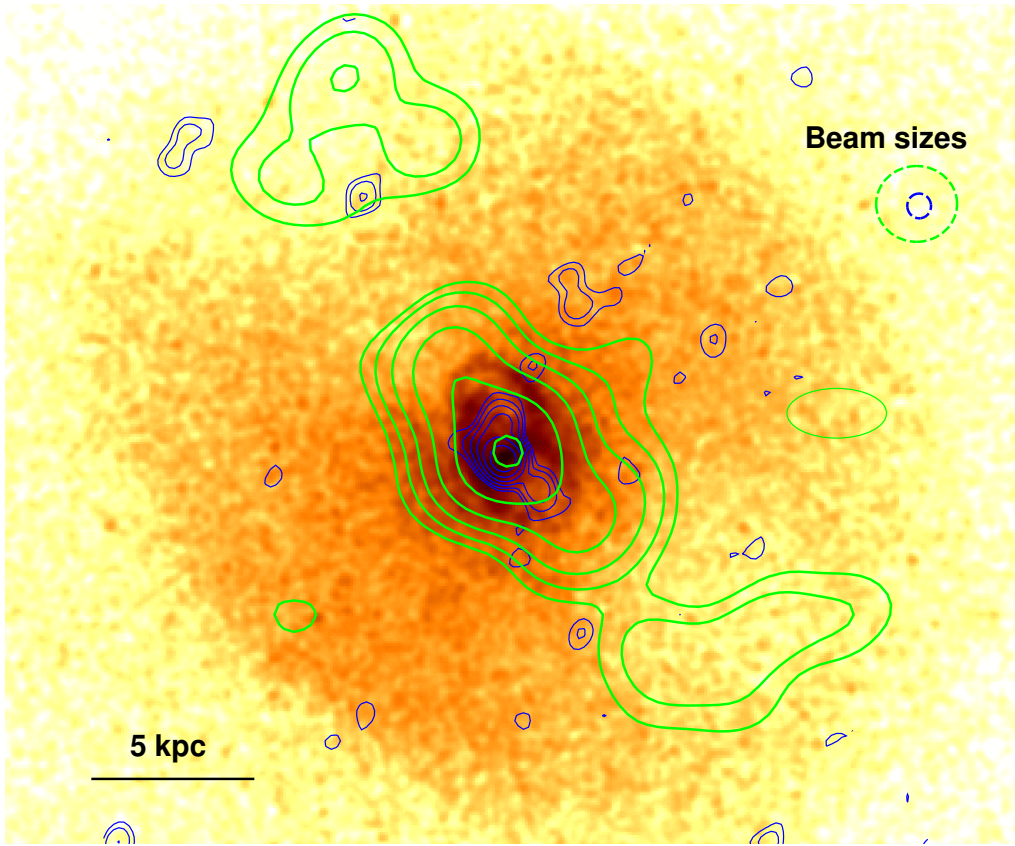}
\caption{{\it Chandra} 0.3--2.0~keV image of the core, with 1.36~GHz
  {\it VLA} B array configuration (blue) and 235~MHz {\it GMRT} (green) contours
  overlaid.  
  The contours start at 3$\sigma$ and are spaced by a factor of two.
  Low frequency radio emission fills the intermediate cavities near
  the edge of the FOV, and emission at both frequencies fills the
  inner cavities. 
  The central peak of the 1.36~GHz contours is coincident with the
  AGN, while the peak of the 235~MHz contours is northwest of the AGN,
  roughly coincident with the center point of the elliptical edge
  defined by the inner shock front.
  The dashed lines indicate
  the beam sizes for each radio observation.
  \label{fig:core_img}
}
\end{figure}

\begin{figure}
\plotone{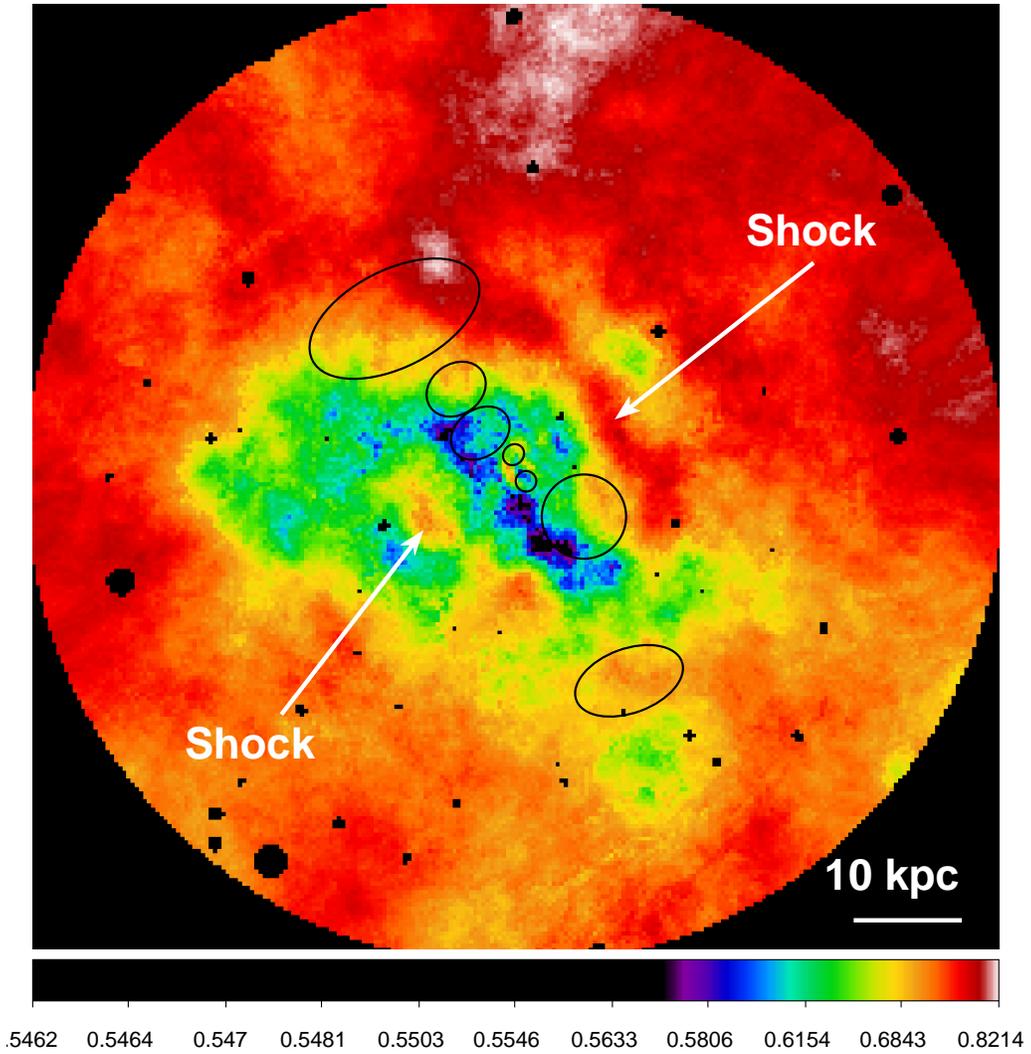}
\caption{Temperature map derived from the ACIS-S3 data, with the
  cavity regions from Figure~\ref{fig:cavoverlay} overlaid.
  The color-bar gives the temperature in keV.  The shocks
  stand out as temperature peaks coinciding with the surface brightness
  edges.  Cooler gas follows the
  southeastern edges of the intermediate surface brightness cavities
  indicated in Figure~\ref{fig:fullimg}.
  \label{fig:tmap}
}
\end{figure}

\begin{figure}
\plottwo{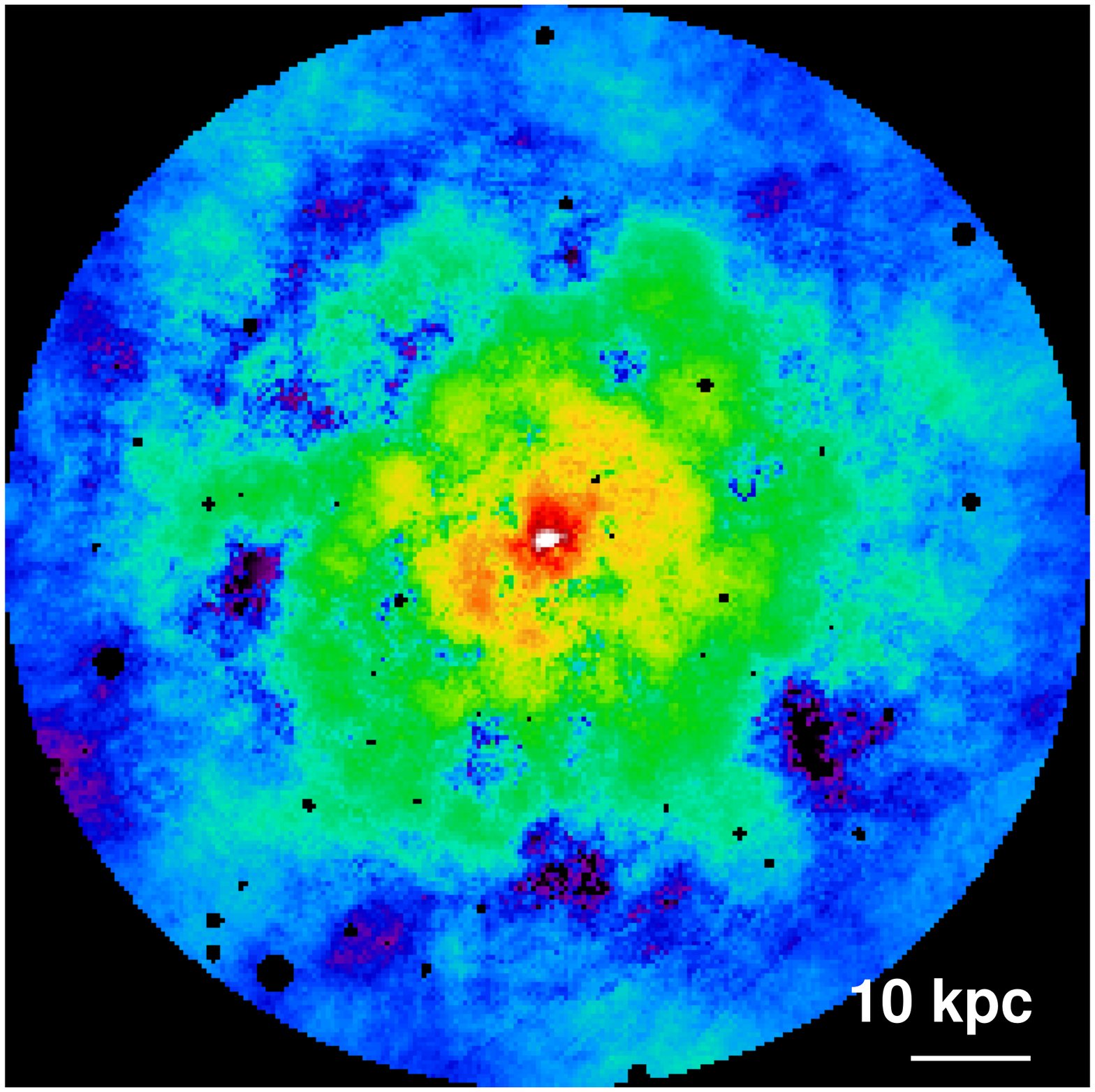}{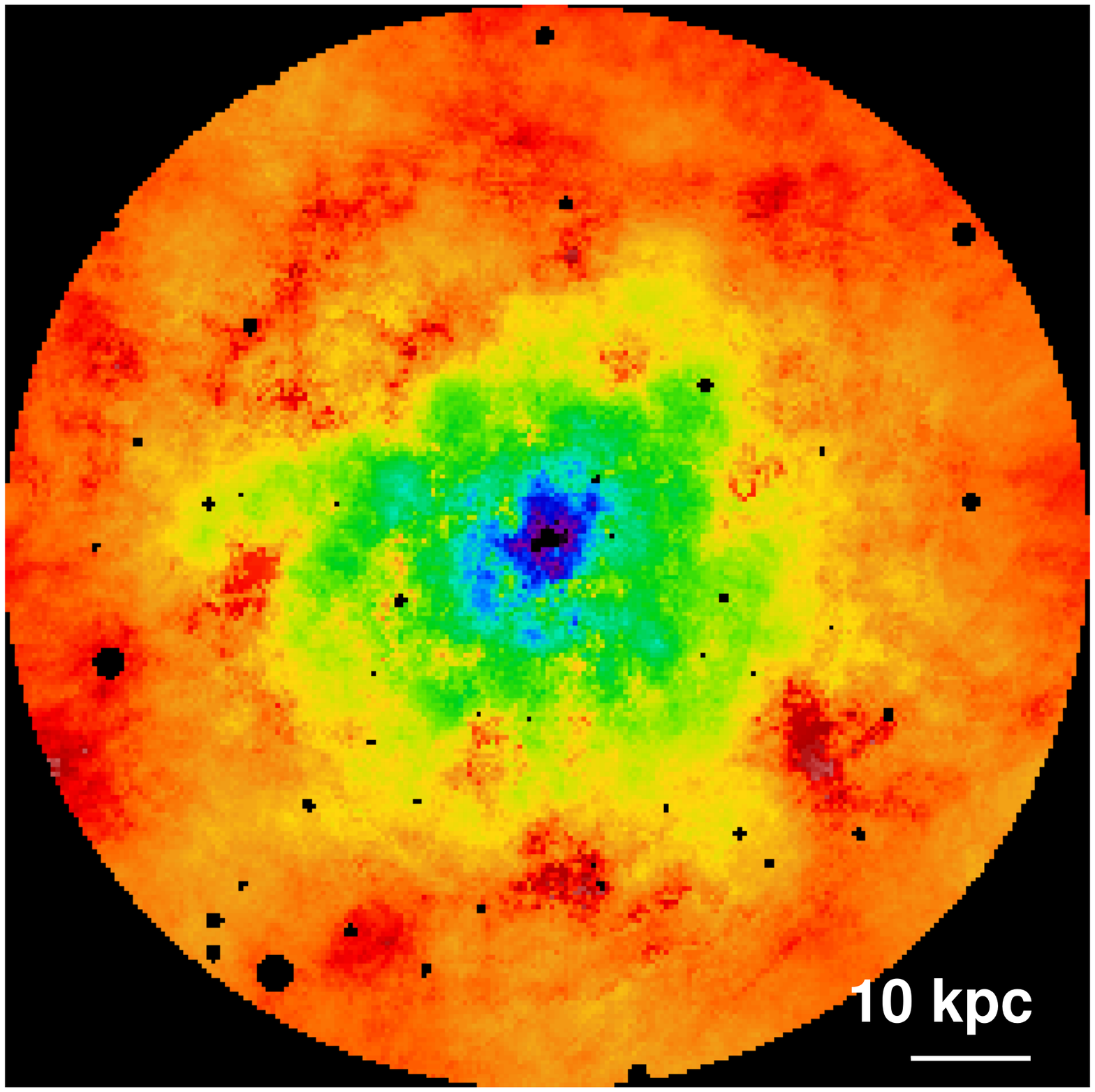}
\caption{
  Pseudo pressure (left) and entropy (right) maps, in arbitrary units.
  The pressure map was calculated as $kT A^{1/2}$ and the entropy map
  as $kT A^{-1/3}$, where $A$ is the {\sc apec} normalization scaled
  by the area of the extraction region.  The pressure jumps at the
  10~kpc shocks are visible $\sim 10$~kpc northwest and southeast of
  the central peak.
  \label{fig:press}
}
\end{figure}

\begin{figure}
\plotone{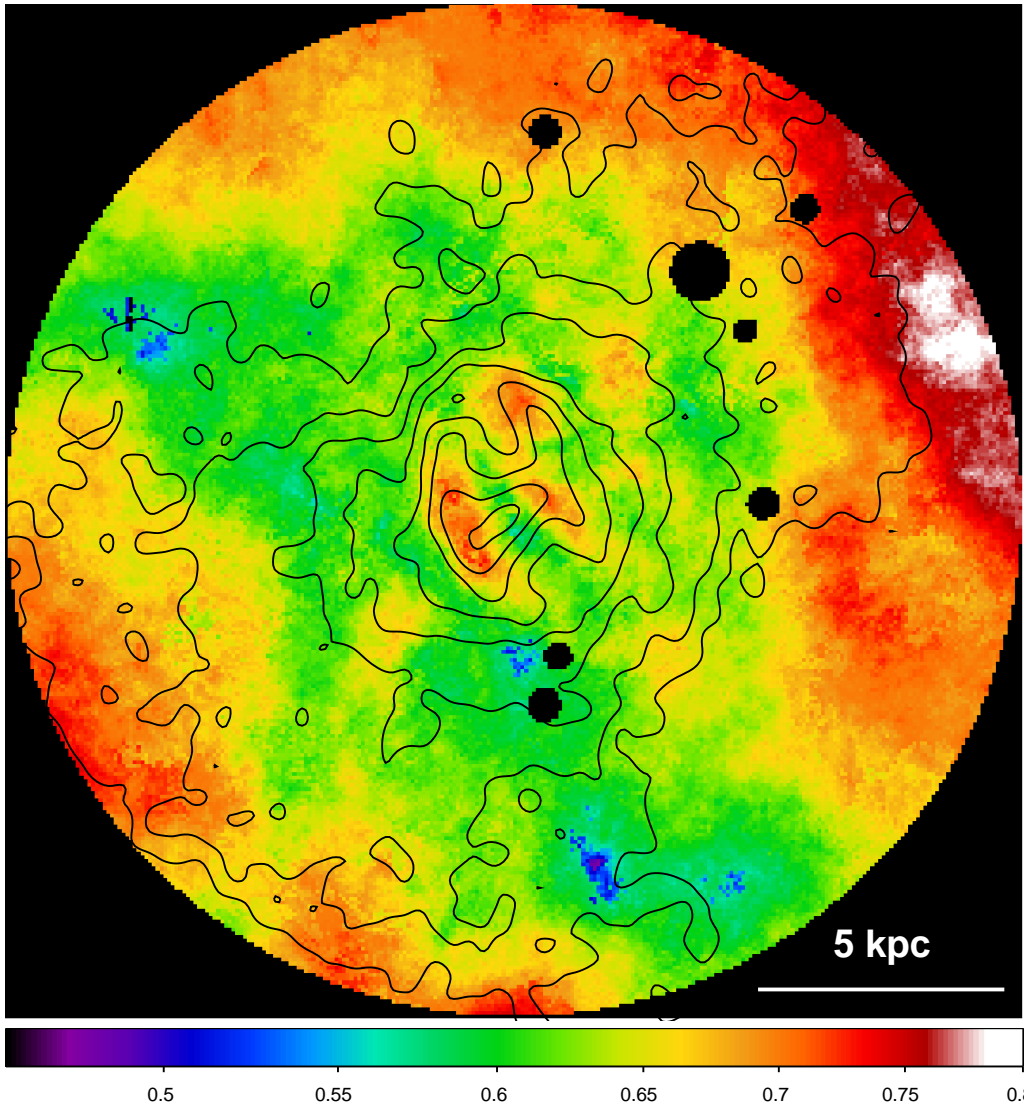}
\caption{A high-spatial resolution temperature map of the core of
  NGC~5813, with 
  {\it Chandra} X-ray 0.3--2.0~keV logarithmic surface brightness contours
  overlaid. The color-bar gives the temperature in keV. The bright
  rims surrounding the innermost cavities are revealed to contain hot
  gas, probably shock-heated by the central AGN during a recent outburst.
  \label{fig:tmap_core}
}
\end{figure}

\begin{figure}
\plottwo{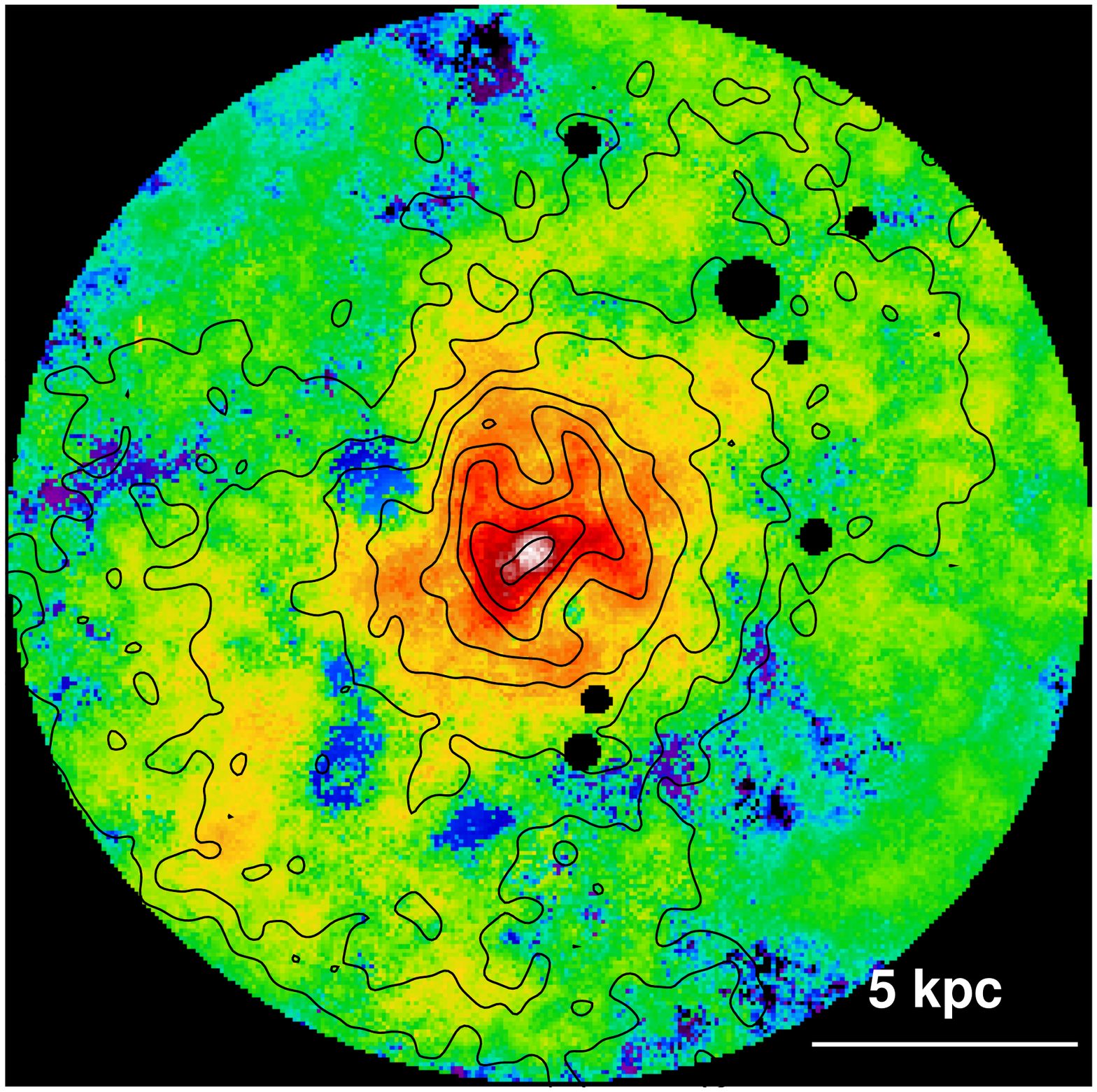}{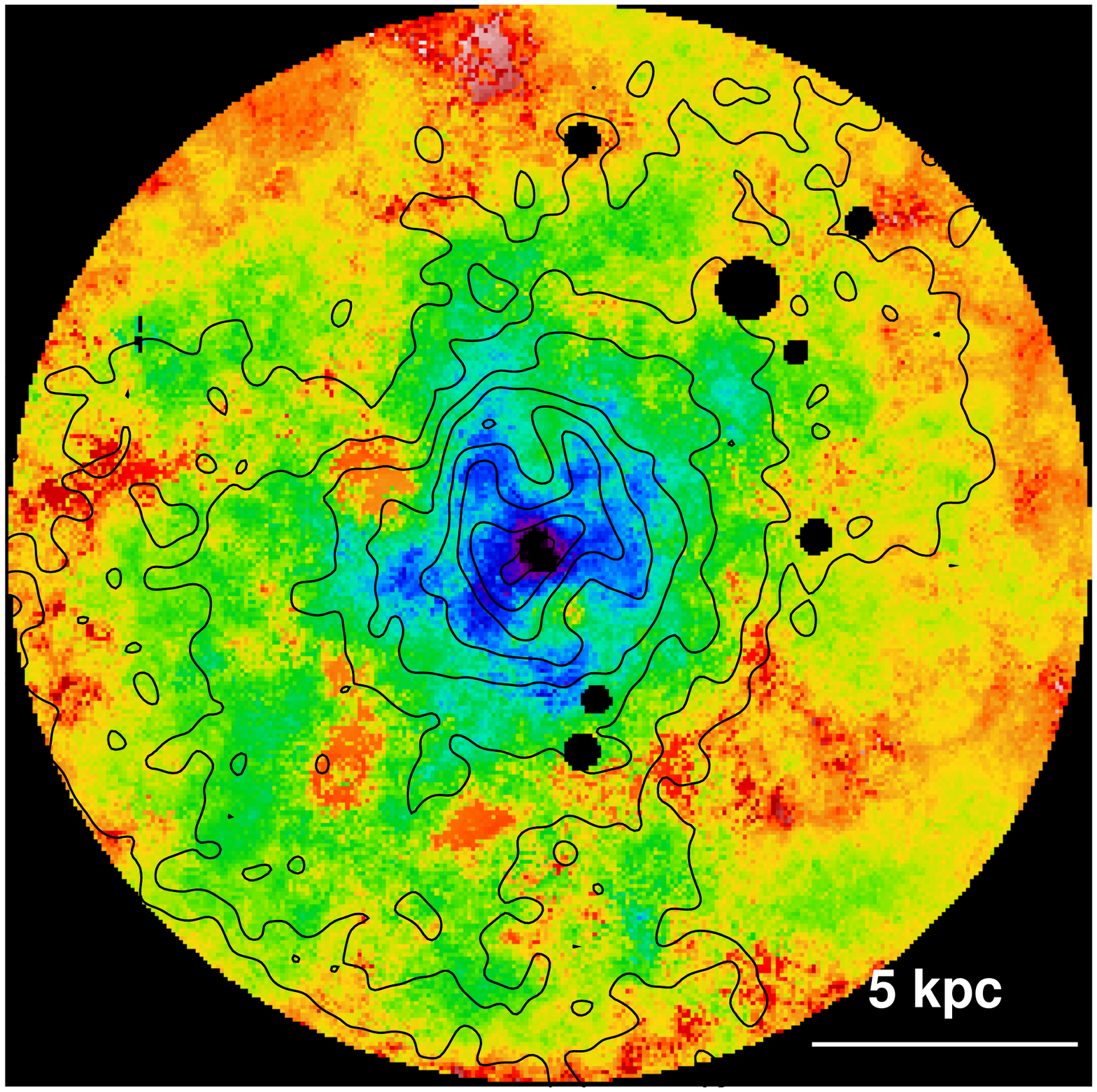}
\caption{
  Pseudo pressure (left) and entropy (right) maps, in arbitrary units,
  derived as in Figure~\ref{fig:press}.  X-ray surface brightness
  contours are overlaid in black.
  There is a sharp pressure jump coincident with the temperature jump
  southeast of the core seen in Figure~\ref{fig:tmap_core}, identifying
  this feature as a shock.  Also visible are local pressure minima
  (and entropy maxima) at the location of the inner cavities,
  presumably due to the additional non-thermal pressure support from the radio
  emitting plasma in the cavities.
  \label{fig:core_press}
}
\end{figure}

\begin{figure}
\plotone{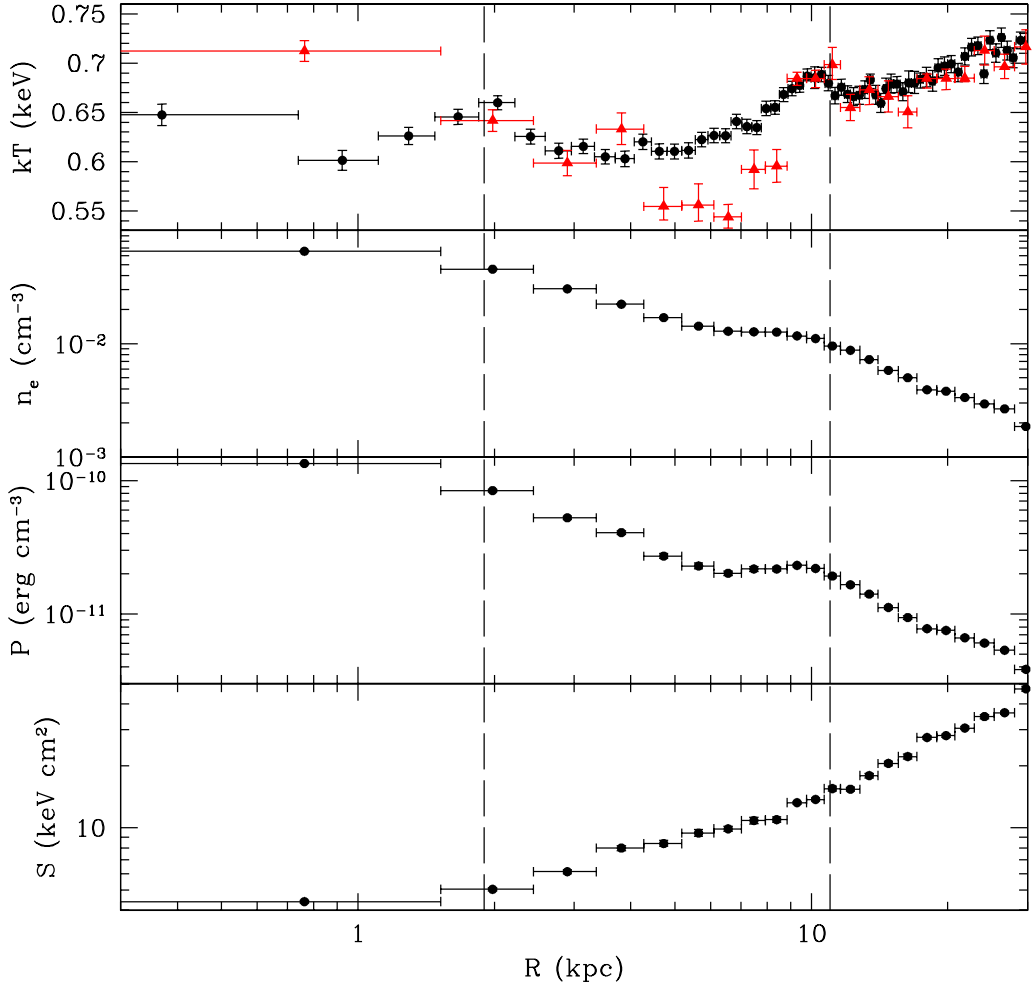}
\caption{
  Azimuthally averaged radial profiles for (top to bottom)
  temperature, electron density, pressure, and entropy, extracted from
  circular annuli.  The top
  panel shows the projected (black circles) and deprojected (red
  triangles) temperature profiles.  All other panels show deprojected
  values.  The vertical dashed lines mark the average radius of the
  1.5~kpc (left) and 10~kpc (right) elliptical shock fronts indicated
  in Figure~\ref{fig:fullimg}.
  The shocks are
  smeared over multiple bins due to their non-spherical structure.
  The temperature and pressure increase at each shock front.
  The temperature profiles show a
  central temperature spike, even though the region of the central AGN
  has been excluded from the fits.  This is because the profile center
  is in the region of the hot 
  overlapping rims of the central cavities, shown in Figure~\ref{fig:xcore}.
  The dips in pressure and density at $\sim$6~kpc are due to the X-ray
  cavities.  
  \label{fig:azprof}
}
\end{figure}

\begin{figure}
\plottwo{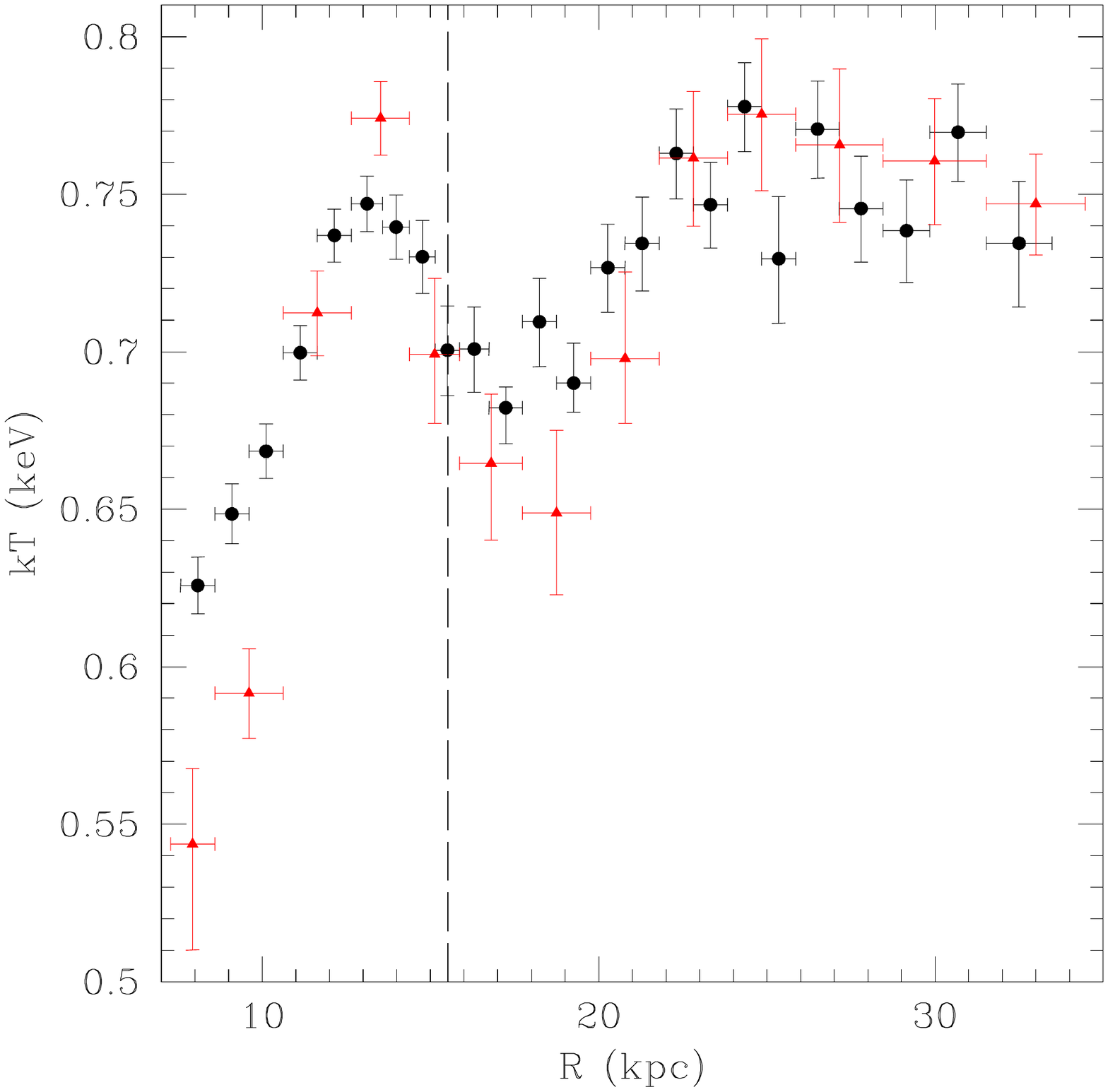}{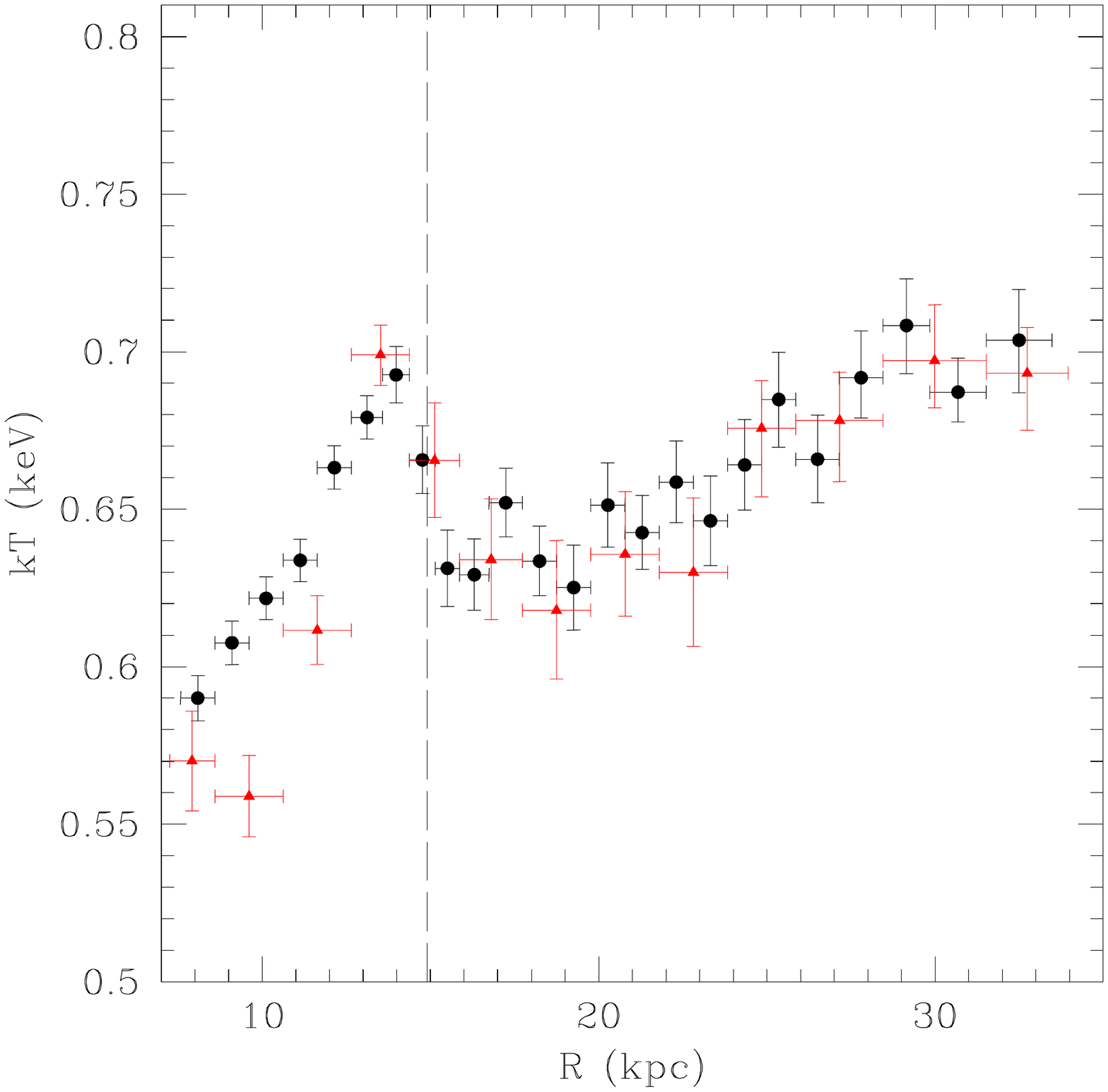}
\caption{
  Projected (black circles) and deprojected (red
  triangles) temperature profiles across the northwestern (left) and
  southeastern (right) shocks.  Distance is measured from the center of
  curvature of the corresponding edge and should not be compared with
  distances in the azimuthally averaged profiles.  The dashed
  lines mark the positions of the density jumps calculated in
  \S~\ref{sec:shocks}.
  \label{fig:sec_ktprof}
}
\end{figure}

\begin{figure}
\plotrtwo{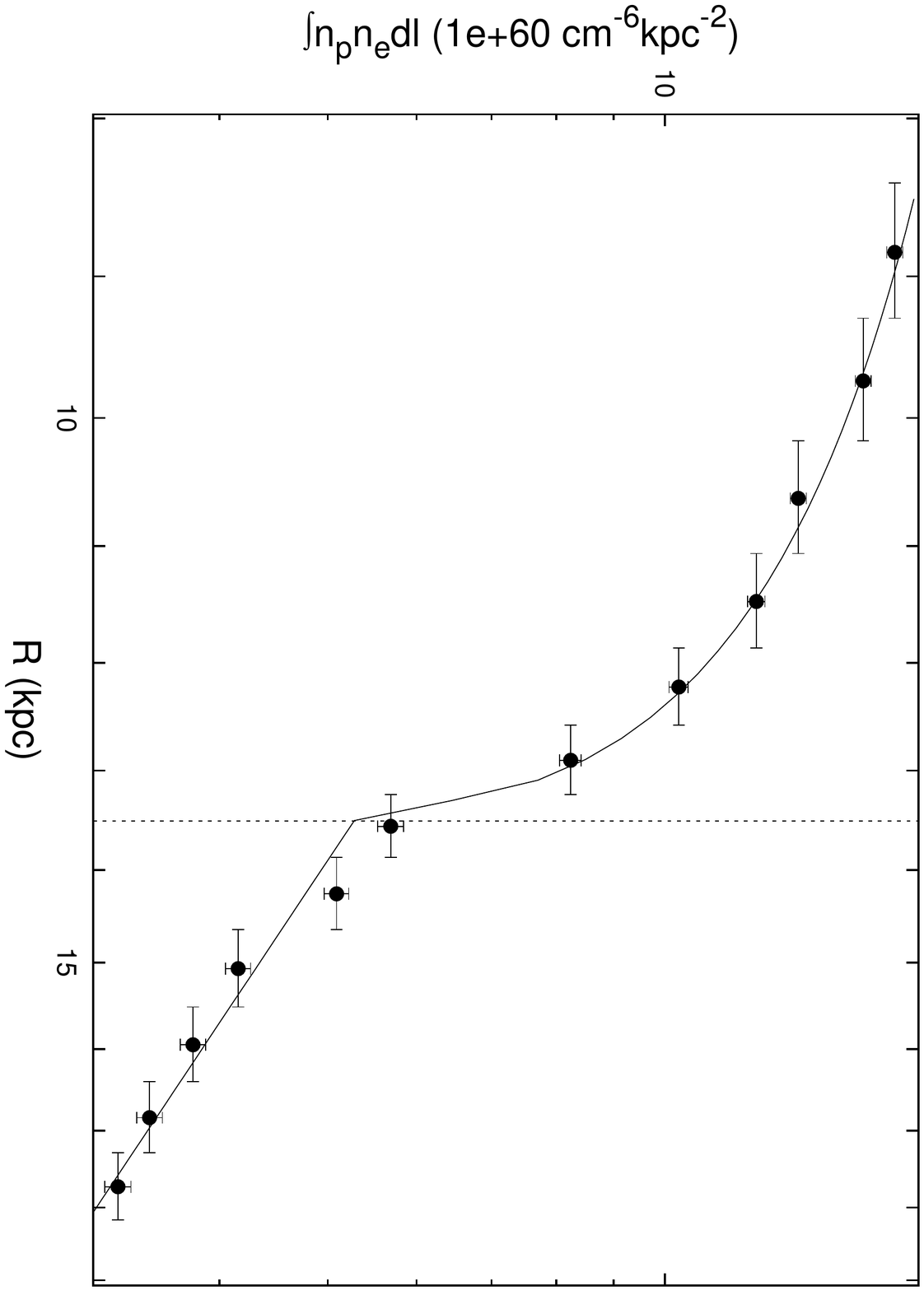}{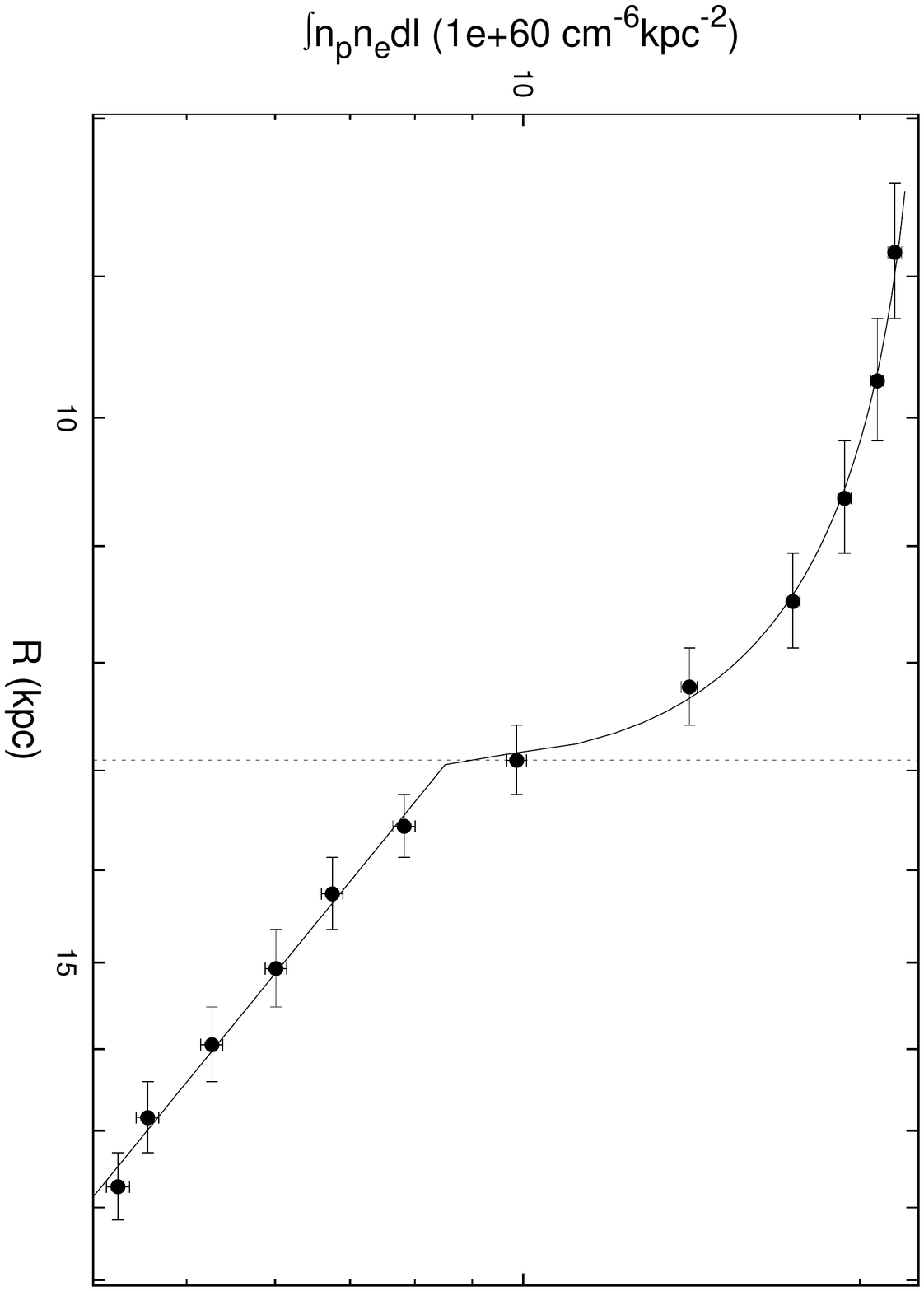}
\caption{
  Integrated emission measure profiles across the northwest (left) and
  southeast (right) edges around the middle cavities, indicated in
  Figure~\ref{fig:fullimg}. 
  Distances are measured from the center of curvature defined by the
  shock, not from the central AGN.
  Each
  profile shows a sharp edge at $\sim$13~kpc.
  The solid lines show the projected best-fitting discontinuous power-law
  density models, and
  the vertical dashed lines mark the best-fitting break radii
  corresponding to density discontinuities.
  \label{fig:edges}
}
\end{figure}

\begin{figure}
\centering
\leavevmode
\includegraphics[angle=90, width=\columnwidth]{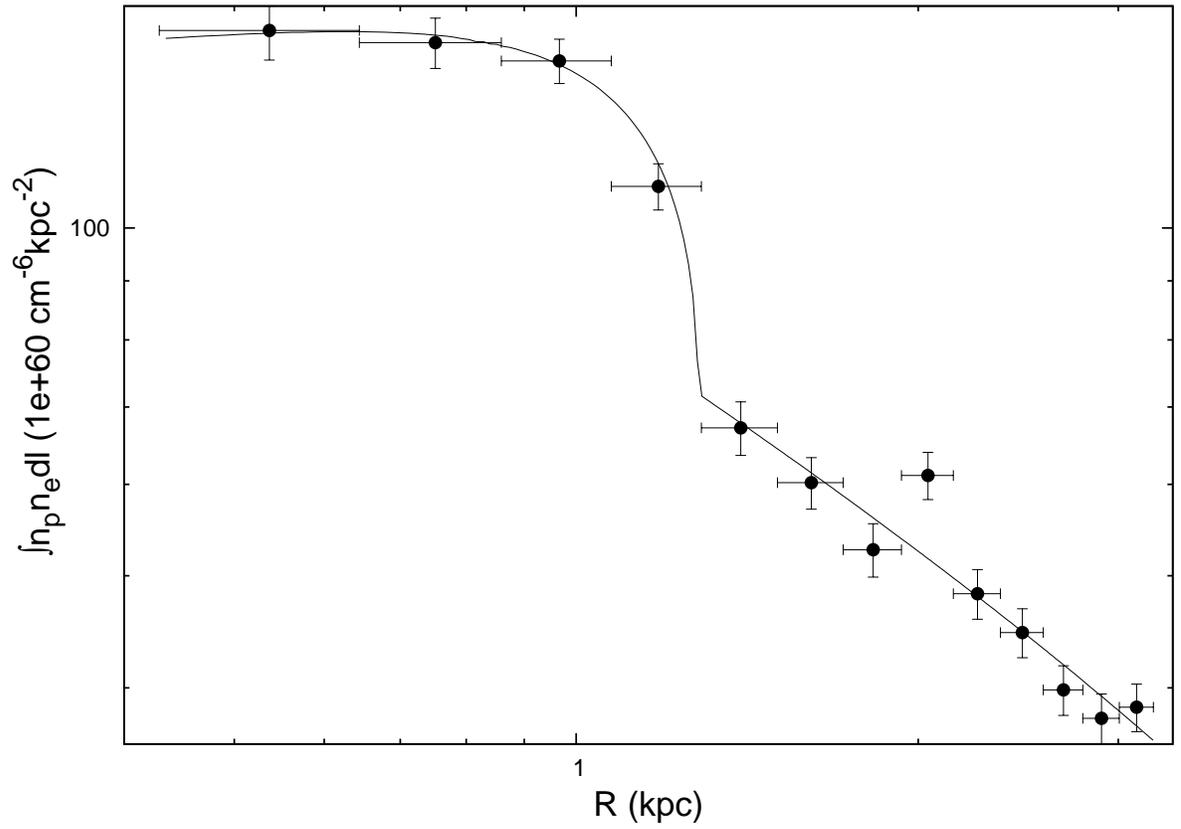}
\caption{
  Integrated emission measure profile across the 1.5~kpc edge just southeast
  of the core, most clearly visible in Figure~\ref{fig:xcore}, which is
  coincident with the edge seen in the core temperature map
  (Figure~\ref{fig:tmap_core}).
  \label{fig:core_emfit}
}
\end{figure}

\begin{figure}
\centering
\leavevmode
\includegraphics[angle=270, width=\columnwidth]{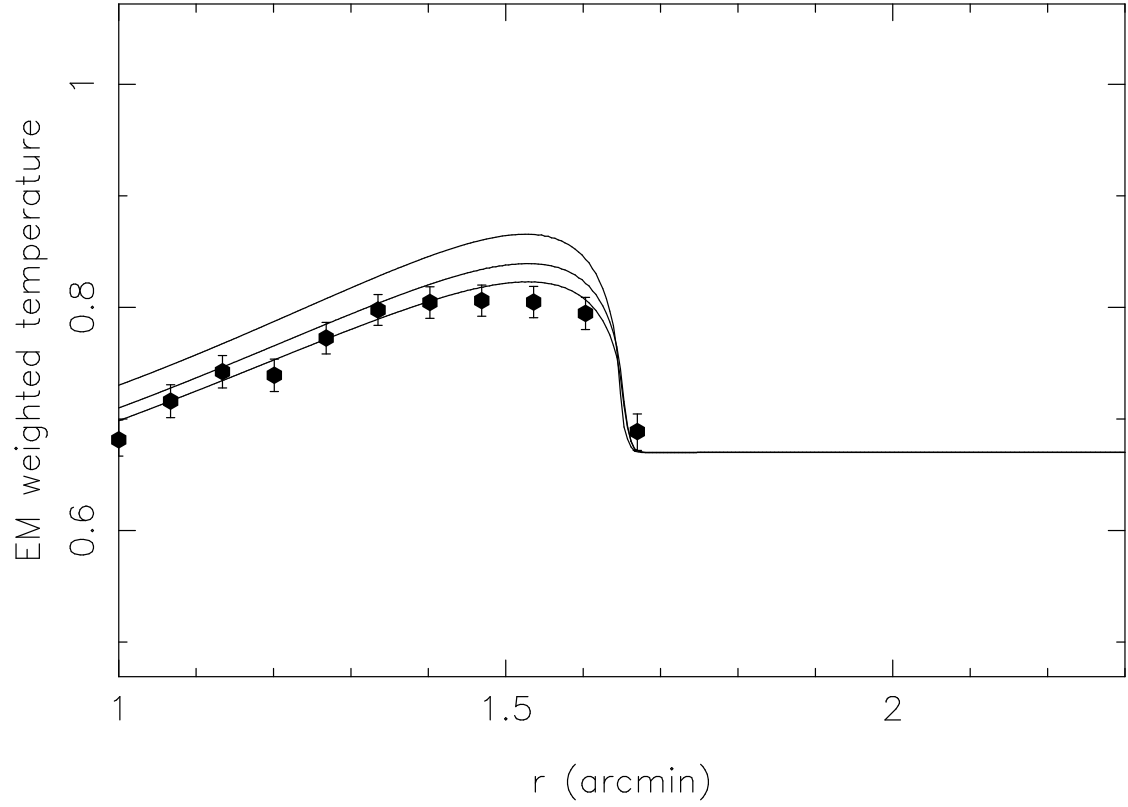}
\caption{
Temperature profile from our hydrodynamical simulations of a point
explosion in an isothermal sphere, chosen at a time when the
corresponding surface brightness profile best matches the observed
profile for the northwestern outer shock.  The solid lines show the
emission-weighted projected temperature profile, with 90\% confidence
intervals.  The points are from results of {\sc xspec} simulations,
with the model folded through the {\it Chandra} response and
projected.  The predicted projected temperature rise is consistent with
what is seen in Figure~\ref{fig:sec_ktprof}, even though the true
temperature rise in the model is larger and consistent with
the prediction from 1D shock jumps conditions based on the Mach number.
\label{fig:hydro_projkt}
}
\end{figure}

\begin{figure}
\plotone{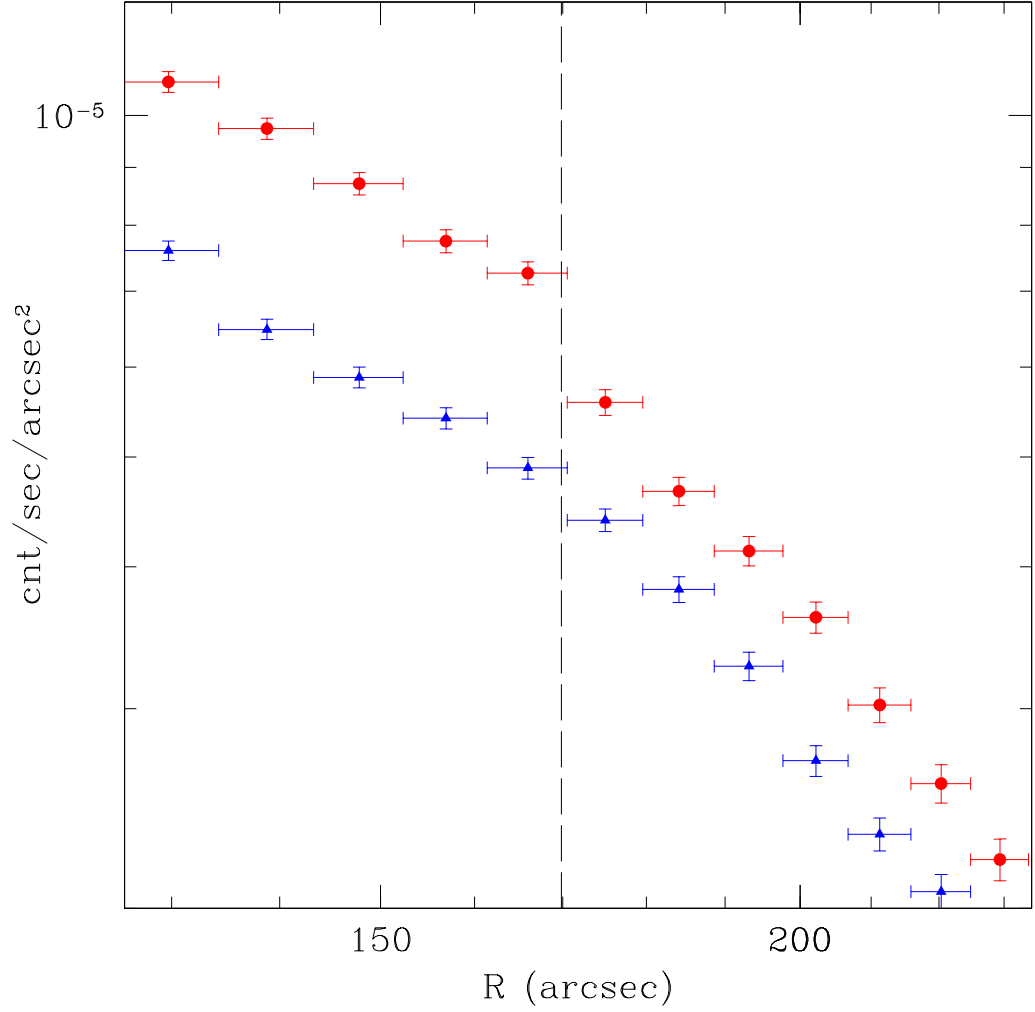}
\caption{
0.3--2.0~keV background-subtracted surface brightness profiles in two
wedges across the outer edge indicated in 
Figure~\ref{fig:oedge} to the northwest (red circles) and south (blue
triangles).  The northwestern profile was measured  between 12\mydeg\
- 79\mydeg\ and the southern profile between 206\mydeg\ - 320\mydeg\
(angles measured from west to north).  The northwestern profile shows
a jump, followed by a change in slope, at $\sim$170\arcsec\ (vertical
dashed line), roughly
the location of the outer edge.  The southern
profile also shows a change in slope at $\sim$170\arcsec, though with no
associated jump.
\label{fig:oprof}
}
\end{figure}

There is a weak surface
brightness edge (most clearly seen in Figure~\ref{fig:oedge})
surrounding most of NGC~5813 at a distance of $\sim$160\arcsec\
(25~kpc) from the central AGN.  
To check the significance of this feature we
extracted the surface brightness profile across this edge in two
sectors, one to the northwest between 12\mydeg\ - 79\mydeg\ (measured
north from west, the same
angular range spanned by the bright section of the elliptical edge to
the northwest) and a wider wedge between 206\mydeg\ - 320\mydeg\ to
the south.  The wedges were chosen to match the curvature of the outer
edge, so that the profiles were off center from the peak of
the overall diffuse emission.  The resulting surface brightness
profiles are shown in Figure~\ref{fig:oprof}.  Both profiles show a
similar 
change in slope at $\sim 170$\arcsec\ (26~kpc), at the position of the outer
edge, and the northwestern profile
shows a sharp jump at the same location.  We conclude  that the outer
edge-like feature indicated in Figure~\ref{fig:oedge} is real, and
corresponds to a change in slope and possibly a discontinuity in the
surface brightness profile.
The edge lies just beyond the outer X-ray cavities
indicated in Figure~\ref{fig:oedge}, and the association is
reminiscent of the 1.5~kpc and 10~kpc shocks to the inner and
intermediate cavities.
The discontinuous jump is also stronger in the northwest
than in the south, consistent with what is seen for the
elliptical 10~kpc shock front edge.
It may therefore represent the weak
remnants of a shock associated with this older outburst.  

To determine the nature of this edge, we fit the surface brightness
profile in the wedge to the the south,  where this feature is the sharpest.
Although there is a significant surface brightness discontinuity to the
northwest (see Figure~\ref{fig:oprof}), we focus on the wedge to the
south, since we have better statistics in the wider southern wedge,
and since the shape of the northwestern discontinuity is not
well-described by our 
shock model density profile (compare the profile shapes in
Figure~\ref{fig:oprof} and in Figure~\ref{fig:edges}).
The shape of the surface brightness profile is obviously not
well-described by a single power-law, and modeling this profile with a
projected power-law density model did not provide an acceptable fit.
We also fit the profile with a $\beta$-model density profile, which
gave a poor fit with  $\chi^2_\nu = 13.9/6$.
The fitted model showed asymmetric residuals in the region of the
surface brightness edge, so we tried fitting the 
edge with a projected discontinuous power-law density model, as in
\S~\ref{sec:shocks}.  This gave an improved fit, with $\chi^2_\nu =
0.4/4$, and a density jump factor of $A_{\rm outer} =
1.06^{+0.12}_{-0.13}$, consistent with no jump.   The inner and outer
slopes of the density profile were 
$\alpha = -0.62^{+0.09}_{-0.18}$ and $\beta = -2.30^{+0.25}_{-0.32}$.
Unfortunately, there were 
inadequate statistics to measure the temperature and abundance profiles
across this outer edge and confirm it as an old shock.
In particular, the edge could in principle be a metallicity
edge, or the interface between the galaxy
atmosphere and the extended group atmosphere (where one would expect
to see a change in slope of the surface brightness profile).  We
conclude that, while 
the outer edge is 
consistent with an old shock with $M \approx 1.1$ associated with the
outermost X-ray 
cavities, the data are also consistent with no density jump and
further observations are 
needed to determine its nature.

\subsection{The X-ray Cavities} \label{sec:cavities}

\begin{figure}
\plottwo{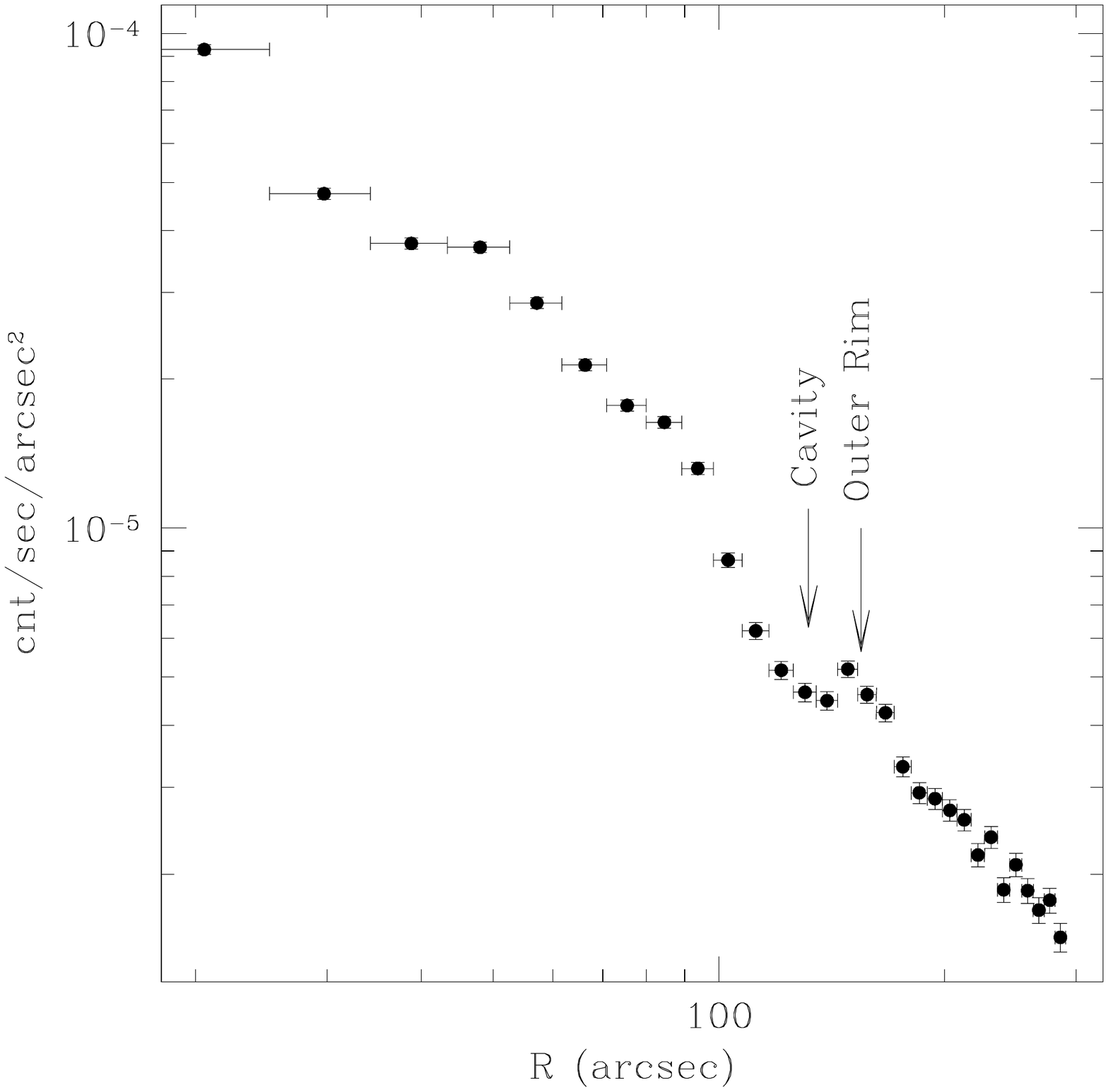}{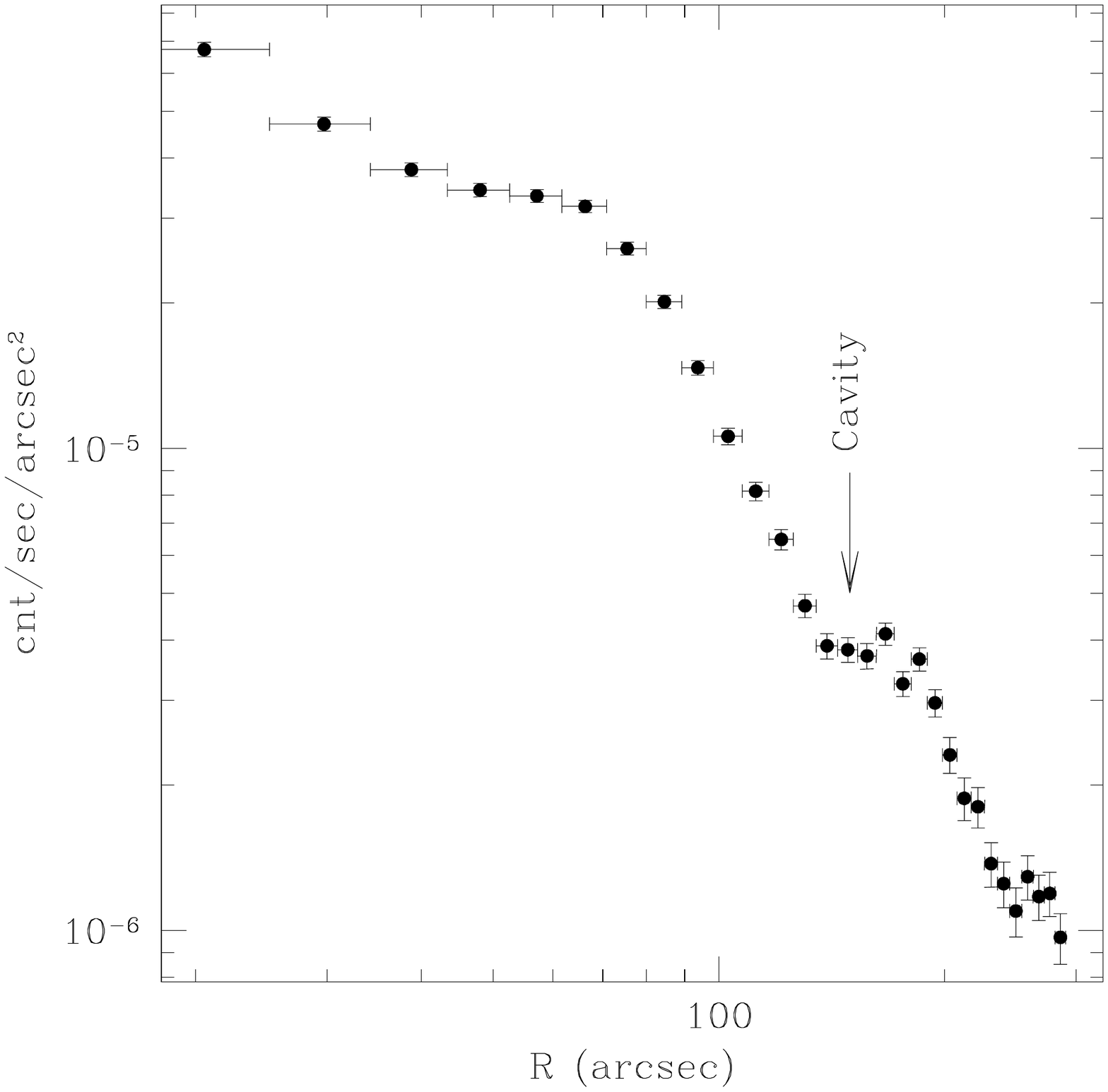}
\caption{
  0.3--2.0~keV background subtracted surface brightness profiles
  across the outer cavities indicated in Figure~\ref{fig:oedge}. 
  For the northeastern cavity (left), the profile was extracted between
  112\mydeg\ - 153\mydeg\ (measured from west to north), and between
  286\mydeg\ - 310\mydeg\ for the southwestern cavity (right).  Each
  profile shows a significant dip at the location of the cavity.  For
  the northeastern cavity, the bright outer rim is also visible.
  \label{fig:ocavs}
}
\end{figure}

The X-ray image shows three pairs of roughly collinear cavities
(Figure~\ref{fig:fullimg}, Figure~\ref{fig:cavoverlay}).  Although
most of the cavities are clear, the outermost cavities (in particular,
the southwestern outer cavity) are weak features.
To check their significance, we extracted surface brightness profiles
across each outer
cavity, centered on the overall diffuse emission.  The resulting
profiles are shown in Figure~\ref{fig:ocavs}. Each
  profile shows a significant dip at the location of the outer cavities.
The profile across the northeastern cavity also shows a significant hump just
beyond the cavity dip, corresponding to the bright outer rim seen in
the X-ray images.  For the southwestern outer cavity, the rise just
outside the cavity is less pronounced, since it lacks a bright rim.
We conclude that the faint outer southwestern cavity indicated in
Figure~\ref{fig:oedge} is likely a real feature and represents a
paired cavity to the outer northeastern cavity, each initially created
by the same AGN outburst from the central SMBH.

Each of the three pairs of X-ray cavities in NGC~5813 is likely
associated with a distinct AGN outburst.  
Although the position angle of the inner and intermediate cavity pairs appears
to vary slightly, by about 10\mydeg\ - 15\mydeg, possibly indicating
that the central black hole that has inflated the cavities is
precessing, the cavities are roughly collinear.
The fact that the cavities
are regular and collinear suggest that they have evolved passively,
i.e., have not been disturbed by gas motions due to sloshing,
turbulence, mergers, etc., consistent with previous conclusions on the
dynamical state of this system (Tran \etal\
2001; Emsellem \etal\ 2007).   Measuring the 
properties of these cavities is therefore of great interest, since a
comparison of the different pairs will provide information on how the
cavities evolve, and on the outburst history of the central AGN (in
particular, we want to know whether the outbursts have a similar total
energy output and
mean power, as
might be expected in a near ``steady state'' AGN feedback model, or whether the
outbursts change significantly even in an otherwise apparently relaxed
system like NGC~5813).

\subsubsection{Cavity Ages and Energies}\label{sec:cavprop}
Table~\ref{tab:cavities} summarizes the properties of the X-ray
cavities, which are shown as overlaid regions in Figure~\ref{fig:cavoverlay}.
We assume that the cavities rise in the plane of the sky, that the
mass of material within the cavities is negligible, and that
each cavity has a spherical or oblate spheroidal geometry with the
minor axis in the plane of the sky.
The former assumption is suggested by the fact that the cavities are
detected in the image, and that the inner two pairs
are near the same projected radii as their associated shock fronts.  If
the global gas distribution was significantly extended along the line
of sight, with the bubbles rising along this extension, they would be
difficult to detect for large inclination angles of the extension due
to the large column of cool gas at smaller radii behind (or in
front of) the bubbles (see B\^{i}rzan \etal\ 2009).
Columns~2~\&~3 in Table~\ref{tab:cavities} give the cavity major and
minor axis, respectively, and Column~4 gives the distance from the AGN
to the center of the cavity.
The rise time of the cavities, given in Column~5, is calculated
assuming that they rise buoyantly at half the sound speed $c_s$,
similar to what 
is found from simulations (\eg, 0.6--0.7$c_s$ in Churazov \etal\
2001).  For a $kT = 0.65$~keV gas, the
sound speed is $c_s = 416$~\kms .  Note that for the innermost
cavities, the distance from the AGN is comparable to the cavity size,
suggesting that they are currently being, or have only recently been, inflated
by the AGN.  Therefore the 
rise times, which are computed using the distance from the AGN to the
cavity center, are not a reliable estimate of the true age, since
early on the cavities are driven by the momentum of the jet (see
B\^{i}rzan \etal\ 2008).

Finally, the mechanical energy required to inflate the bubbles is
given in Column~6, which we estimate as $P V$, where $P$ is the
pressure at the location of the bubble center (taken from the azimuthally
averaged pressure profile shown in Figure~\ref{fig:azprof}) and $V$ is
the cavity volume.  The total internal energy of each cavity is expected to be
a few times the mechanical energy ($\sim 3 P V$ for a relativistic plasma,
see McNamara \& Nulsen 2007).

The total mechanical energy is about the same for the southwestern and
northeastern cavities for both the innermost ($\sim 1.3 \times
10^{55}$~erg) and intermediate ($\sim 1.4 \times 10^{56}$~erg)
pairs (taking the sum of the two intermediate cavities to the
northeast), consistent with each pair having formed in two distinct
AGN
outbursts.  The mechanical energies for the outer cavities differ
significantly from one another ($2.6 \times 10^{56}$~erg versus $6.0
\times 10^{55}$
for the northeastern and southwestern cavities, respectively).  We
suggest three possibilities to explain this
discrepancy.  First, the outer southwestern cavity is only marginally
detected, so its measured properties may be inaccurate.  Second, the
outer cavities may be in the process 
of breaking apart, as suggested by their significantly different
volumes, and if they are not devoid of X-ray emitting gas, then the
mechanical energy will be incorrectly estimated. Finally,
Figure~\ref{fig:fullimg} shows that the intermediate
northeastern cavities may connect to the outer northeastern cavity.  If
this is indeed the case, energy may ``leak'' from the intermediate to
the outer cavity, adding to its internal energy and inflating it,
making it easier to detect 
(there is no such connection in the southwest).

A comparison of the ages of the cavities (approximated by the rise
times given in Table~\ref{tab:cavities}) and the shocks (approximated
by the travel time given in Table~\ref{tab:shocks}) for the ``inner'' and
``middle'' features shows that they
are similar ($\sim 10^7$~yr), consistent with the
interpretation that each
set of features was formed at the same time by the same outburst
event.  The cavity ages are systematically larger than the shock ages
derived from our hydrodynamical simulations (by about a factor of 3).  
This is likely due to the 
cavities being initially driven and inflated at some significant
distance from the
central AGN, only to rise buoyantly after the end of the outburst, so
that regarding the cavities as buoyantly rising bubbles early in their
lives is not accurate.
Thus, using the buoyant rise velocities of the cavities likely
overestimates their ages. 

\subsubsection{Pressure Balance with the ICM}\label{sec:cavbal}
The non-thermal radio pressure in X-ray cavities, under the assumptions of
hydrostatic equilibrium and electron dominated pressure, is commonly
found to be less than the pressure of the
surrounding gas derived from X-ray observations (B\^{i}rzan \etal\ 2008).
To accurately estimate the radio pressure, flux measurements at multiple
frequencies are required to estimate the spectral index  $\alpha_r$ of
the radio emission.  In NGC~5813, only the innermost cavities are
detected at more than one frequency (see \S~\ref{sec:rimg}).
Unfortunately, the large beam size and
contamination from extended emission outside the cavities at
235~MHz make it difficult to accurately measure $\alpha_r$.  
Comparing the total emission in the region of the inner cavities,
including contributions
from extended emission outside of the central cavities at 235~MHz and
1.36~GHz (the latter detected in archival {\it VLA} C array configuration
observations) and emission from the core, gives $\alpha_r = 0.88$.
If we
assume $\alpha_r = 0.7$ within the inner cavities, a typical value for young
radio lobes (B\^{i}rzan \etal\ 2008), we 
find that the non-thermal pressure in each cavity is $P_{rad}
\sim 2.3 \times 10^{-12}$~erg~cm$^{-3}$ (where we have used the
revised equipartition equations of Brunetti \etal\ 1997 with a
low-energy cut-off of $\gamma_{\rm min} = 100$).  This is much less
than the X-ray gas pressure outside the cavities $P_{gas} ~\sim
10^{-10}$~erg~cm$^{-3}$ (see Figure~\ref{fig:azprof}).  Balancing the
pressure with low energy electrons alone would require $\alpha_r \approx 2$,
which is unusually steep for young radio cavities (B\^{i}rzan \etal\
2008).  If we assume $\alpha_r = 0.7$ and vary the ratio of the
energy in protons to the energy in electrons $k$, we find that $k
\approx 2000$ is required to balance the non-thermal and X-ray gas
pressures, which is typical for radio galaxies in cool cores (B\^{i}rzan \etal\
2008; Dunn \etal\ 2010; Gitti \etal\ 2010).

We compared the radio and cavity power of the inner cavities with predictions
based on the  B\^{i}rzan \etal\ (2008) sample.
The 1.36~GHz radio power for each of the inner cavities is $P_{1360} \approx 1
\times 10^{20}$~W~Hz$^{-1}$.  The relation of B\^{i}rzan \etal\ (2008)
then predicts a total cavity power of $P_{\rm cav} \approx 9 \times
10^{42}$~erg~s$^{-1}$.  
Taking the cavity ages and mechanical energy ($P V$) from
Table~\ref{tab:cavities} and assuming total cavity energies of $3 P V$
gives a power of $P_{\rm cav} \approx 4 \times 10^{41}$~erg~s$^{-1}$
for the inner cavities, more than an order of magnitude less than the
estimate from B\^{i}rzan \etal\ (2008).
However, it is within the range of the large scatter in the B\^{i}rzan
\etal\ (2008) sample (see their Figure~6).
More recently, Cavagnolo \etal\ (2010) derive a relation between
$P_{\rm cav}$ and $P_{\rm radio}$ for lower mass systems and find a
steeper slope than B\^{i}rzan \etal\ (2008).  Their sample includes
results from the {\it Chandra} observations of NGC~5813 we consider
here, so that NGC~5813 is consistent with their derived relation.


\subsection{Buoyantly Lifted Gas} \label{sec:uplift}

\begin{figure}
\plotone{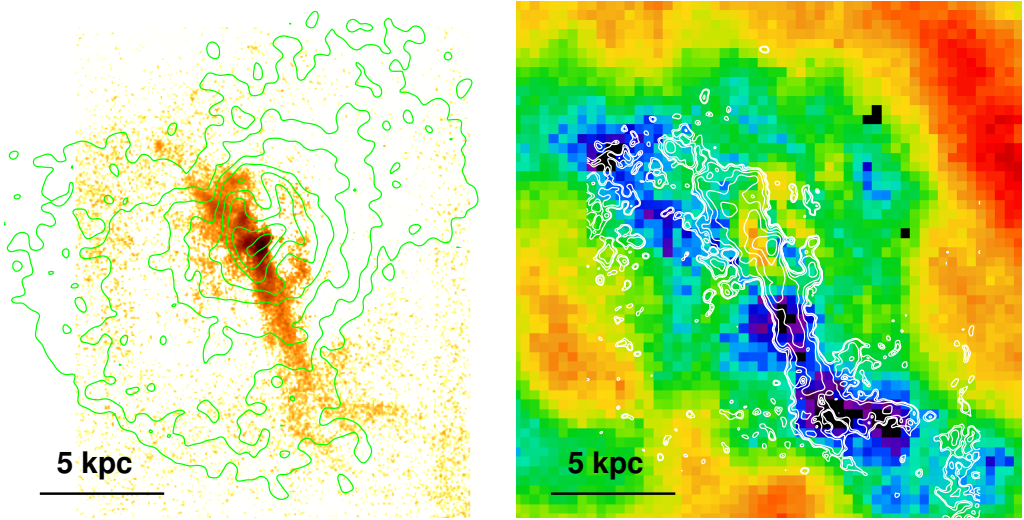}
\caption{
  {\it Left:} H$\alpha$ image of the core of NGC~5813 taken with the {\it SOAR}
  telescope.  The {\it Chandra} X-ray contours are overlaid in green.
  The H$\alpha$ filaments extend to the southern edges of the
  intermediate cavities, and are co-spatial with the trail of cool gas
  seen in the X-ray temperature map in Figure~\ref{fig:tmap}.  In the
  center, the H$\alpha$ emission anti-correlates with the inner X-ray
  cavities.  The image shows the cool H$\alpha$ gas being displaced as
  the central cavities are inflated, and lifted by the intermediate
  cavities as they rise buoyantly. {\it Right:} Close-up of the
  central part of the temperature map shown in Figure~\ref{fig:tmap},
  with H$\alpha$ contours overlaid in white.  The H$\alpha$ is
  co-spatial with the cool X-ray filament.
  \label{fig:halpha}
}
\end{figure}

The temperature map (Figure~\ref{fig:tmap}) shows an extension of cool
gas along the line defined by the X-ray cavities, offset $\sim 3$~kpc to
the southeast.  This cool gas extends 11~kpc to the inner edges of the
intermediate cavities.  The most natural interpretation of this
feature is cool gas that has been buoyantly lifted by the intermediate
X-ray cavities, as seen in other systems (Forman \etal\ 2007; Simionescu \etal\ 2008;
Kirkpatrick \etal\ 2009; Simionescu \etal\ 2009; 
Randall \etal\ 2010; Werner \etal\ 2010).
The H$\alpha$ image, taken with the {\it SOAR} 
telescope (Donahue \etal\ 2007), with the X-ray surface brightness
contours overlaid, as well as the temperature map with the H$\alpha$
contours overlaid, are shown in Figure~\ref{fig:halpha}.
The H$\alpha$
filaments are co-spatial with the trail of cool gas seen in the X-ray
temperature map, as seen in many other systems (\eg, Sanders \etal\
2007, 2009), confirming the presence of cool gas.
Assuming line ratios of N[II]$\lambda$6583/H$\alpha$ = 1 and
N[II]$\lambda$6548/H$\alpha$ = 0.35, the total H$\alpha$ flux is
$F_{{\rm H}\alpha} = 9.2 \times 10^{-14}$~erg~s$^{-1}$~cm$^{-2}$ (assuming no intrinsic
absorption).  The Kennicutt relation
for the star formation rate
\begin{equation}
{\rm SFR} (M_\odot/{\rm yr}) = 7.9 \times 10^{-42} L_{{\rm H}\alpha} ({\rm erg~s}^{-1})
\end{equation}
(Kennicutt 1998) then gives SFR $= 0.09 M_\odot$~yr$^{-1}$.  
This assumes that the H$\alpha$ emission is completely driven
by UV radiation from young stars, which may not be the case 
(\eg, Ferland \etal\ 2009), and that there is no intrinsic absorption.
A violation of the former assumption would give an overestimate of the SFR,
while a violation of the latter would give an underestimate.
We compared the derived SFR to the 
mass cooling rate estimated from the X-ray observations by fitting the spectrum of
the cool filament with a {\sc vapec} plus a cooling flow {\sc vmcflow} model, with the
abundances fixed at the best fitting values for the total diffuse
emission given in \S~\ref{sec:total}.  This gives a mass cooling rate in
the gas of $0.41 \pm 0.07 \, M_\odot$~yr$^{-1}$.  
One concern with this result is that fitting a multi-temperature gas
with a single cooling flow component may boost the inferred mass
accretion rate.  Unfortunately, we were unable to obtain a fit with
reasonable parameter constraints from the {\it Chandra} data using a
model combining an
{\sc apec} and two {\sc vmcflow} components.  We therefore consider the
mass accretion rate inferred from the fit to {\it XMM-Newton} RGS data
described in \S~\ref{sec:total}.  From the two cooling flow
components, we infer upper limits on the mass accretion rate of $< 0.45
\, M_\odot$ ~yr$^{-1}$ and $< 0.25 \, M_\odot$ ~yr$^{-1}$ for gas
cooling above and below 0.4~keV, respectively (these are 2$\sigma$
upper limits).
Therefore, we conclude that the upper limit on the mass accretion
rate from X-ray observations ($< 0.25 \, M_\odot$ ~yr$^{-1}$) and the
star formation rate implied by H$\alpha$ 
observations (SFR $= 0.09 M_\odot$~yr$^{-1}$) are
consistent with star formation being fueled by gas cooling down from
X-ray temperatures.

For the cool gas to be buoyantly lifted by the X-ray cavities,
its total mass must be less than the mass of gas displaced by the
cavities.  
In particular, simulations indicate that the mass of gas buoyantly
lifted by an AGN-blown bubble is about half the mass displaced by the
bubble (Pope \etal\ 2010).
We estimated the gas mass in the southern filament by
fitting the spectra in this region with an {\sc apec} model and
assuming that the cool gas is contained in a cylinder of radius
3.5~kpc and length 9.2~kpc, with the axis in the plane of the sky.
The emission is dominated by the cool gas in this region, and
accounting for the projected hot gas did not significantly change our
results.  We find a total gas mass of $M_{fil} = 1.5 \times 10^8 M_\odot$.
A similar fit to an annulus surrounding the southwestern intermediate
cavity gives an average electron density of $n_e = 0.022$~cm$^{-3}$,
giving a total mass of displaced gas $M_{disp} = 1.6 \times 10^8
M_\odot$, similar to the mass of gas in the filament.  
Thus, $M_{fil}$ is larger than the value predicted by simulations
($\sim 0.5 M_{disp}$) by a factor of two.
We note that deep observations of the buoyantly lifted
filaments in M87 reveal that they have a fine filamentary structure
(Forman \etal\ 2007), in contrast with the solid cylindrical geometry
we have assumed above (Werner \etal\ 2010 argue that the
filling factor in M87 is of order unity in most regions, although they
find a filling factor less than unity in some regions where
there is fine filamentary structure).  
If the filling factor is less than unity then
$M_{fil}$ will be smaller by the same fraction.
We conclude
that the gas mass of the cool filament to the south is consistent with
having been buoyantly lifted by the intermediate southwestern cavity,
and is consistent with simulations
if the filling factor is on the order of $\sim 0.5$.  We also note
that if the filament has indeed been buoyantly lifted by the
intermediate cavity then the filament, and hence the trajectory of
the intermediate cavity, cannot lie far from the plane of the sky
without assuming a small filling factor, since the length of the
filament (and hence the volume it occupies) grows with inclination angle.

\subsection{The Offset of the Central AGN}\label{sec:offset}

As noted in \S~\ref{sec:ximg}, the central AGN is offset $~\sim 0.5$~kpc
southeast of the line defined by the X-ray cavities (see
Figure~\ref{fig:xcore}).  It is also offset from the center points of
the elliptical edges defined by both the 1.5~kpc and 10~kpc shock
fronts (which are each roughly coincident with the line defined by
their respective cavity pairs), by about 2.3\arcsec\ (400~pc) for the
1.5~kpc shock and  
 7.5\arcsec\ (1.15~kpc) for the 10~kpc shock.
A comparison of the {\it Chandra} image with the optical {\it Sloan
  Digital Sky Survey} image (Adelman-McCarthy \etal\ 2008) shows that the
centroid of the optical emission is coincident with the AGN, and is
separated from the center points of the elliptical X-ray shock fronts.  
This suggests that the cD galaxy has some peculiar velocity relative
to the ICM, and that the AGN has moved since first inflating the X-ray
cavities. 
The initial outburst that inflated the inner
cavities occurred about $3 \times 10^6$~yr ago (see
Table~\ref{tab:shocks}).  Since then, the AGN has 
traveled about 400~pc in projection.  This requires a relative
velocity between the galaxy and the ICM of at least $\sim130$~\kms
(this is a lower 
limit since we measure the projected velocity).  Similarly, the
10~kpc shock implies a relative velocity of $\sim100$~\kms.
This is smaller than the host group's (NGC~5846) velocity dispersion (322~\kms,
Mahdavi \etal\ 2005), and the relative radial velocity between
NGC~5813 and NGC~5846 ($\sim 260$~\kms), and is therefore a reasonable
peculiar velocity
for the cD galaxy relative to the group mean (or for the flow velocity
of the ICM gas).  We suggest that during
the last $\sim2 \times 10^7$~yr there has been a relative motion
between the NGC~5813 galaxy and its ICM,  from northwest to southeast,
in projection, 
with the central AGN driving outbursts (\ie, inflating cavities into
the ICM) at two locations (the outer cavities and edge are too poorly
resolved to accurately measure the location of the associated
outburst).  The relative velocity may be due to the peculiar velocity
of the cD relative to the group mean, or to bulk gas motions or gas
sloshing of the ICM, or both.  We note that once the outburst begins and
the cavities are initially inflated, the location at which the energy
is injected into the cavities is unimportant.  The AGN may continue to
move off of the axis of symmetry of a pair of cavities while still
depositing energy into them, driving their expansion and the
resulting shocks.

\subsection{Outburst Energy}\label{sec:energy}

During an AGN outburst, the jets inflate cavities in the ICM, which
do work on the surrounding gas and drive shocks.
It is convenient to regard the work done by the expanding lobes as
``shock energy'', in which case
the outburst energy that is available to heat the ICM is deposited to
the ICM in two forms: the internal energy of the X-ray cavities and
the shock energy.  Since
NGC~5813 shows both cavities and shocks from two distinct
outbursts, we can compare the total energy, mean power, and
the energy budget between 
shocks and cavities for each outburst.  The shock energy can be
roughly estimated 
from the pressure increase across the shock front.  If a total energy
$E$ is added to a gas of volume $V$ the pressure increase is roughly
$\Delta P \sim E/V$.  For a shock with a known Mach number, the ratio
of the post- and pre-shock pressure $f_P = ( P + \Delta P )/P$ is
given by the Rankine-Hugoniot shock jump conditions.  The total shock
energy is therefore 
\begin{equation}\label{eq:eshock}
E \approx P V (f_P - 1).
\end{equation}  
As a consistency check, we compare the estimated shock age, total energy, and
mean power with results from our hydrodynamical model in Table~\ref{tab:shocks}.
The shock age $t_{\rm age,est}$ is estimated as the travel time from the
current position of the shock front to the center point of the
elliptical shock edge, 
assuming a constant Mach number and a sound speed of $c_s = 416$~\kms\
for a $kT = 0.65$~keV gas.  The shock energy $E_{\rm
  sh,est}$ is estimated using equation~\ref{eq:eshock}, assuming a
prolate ellipsoidal geometry for the volume contained within the shock
front with semi-major and -minor axes of 1.44\arcmin\ (13.5~kpc) and
1.13\arcmin\ (10.5~kpc) for the 10~kpc shocks and 17.6\arcsec\ (2.7~kpc) and
11.7\arcsec\ (1.8~kpc) for the 1.5~kpc shocks.  Pressures were taken
from the azimuthal pressure profile shown in Figure~\ref{fig:azprof}.
Although the energies we give from the hydrodynamical model represent
the total outburst 
energy, the point explosion model minimizes the internal energy in
cavities, such that the energy in the central cavity is only a few
percent of the total outburst energy.  The model energy is therefore a very
good approximation of the shock energy, and we refer to it as such.
 The model shock energies
$E_{\rm sh,model}$ were scaled to account for the
difference in total volume for the spherically symmetric model as
compared to the observed elliptical edges, and for the lower average
pressure along the elliptical shock fronts, which cover a range in
radii, assuming the shock energy scales as in equation~\ref{eq:eshock}
(these effects somewhat balance each other, as the volume 
correction increases the total energy, while the pressure correction
decreases it).  The correction factors are between 15--40\%.
The estimated and model shock energies and mean powers agree
reasonably well, within factors of a few, demonstrating the consistency
between rough estimates and results from our point explosion
hydrodynamical model.

Table~\ref{tab:shocks} indicates that the shock energy for the current
outburst is more than an order of magnitude smaller than for the previous
outburst.  The mean power of the current outburst is also
less than that of the previous outburst, by about a factor of
six (1.5$\times 10^{42}$~erg~s$^{-1}$ versus 1.0$\times
10^{43}$~erg~s$^{-1}$, where we take the mean outburst power to be the
sum of the shock energy and the $3 PV$ internal energy of the cavities
divided by the shock model ages).  The lower 
shock energy of the most recent outburst may indicate that
it is ongoing, having only deposited
$\sim$1/40 of its expected total energy output into the observed
shocks (assuming that the current outburst is similar to the previous one).
However, the lower shock energy may simply be a result of the lower mean
power of the current outburst.  We note that from X-ray observations
of elliptical galaxies Allen \etal\ (2006) find
evidence that accretion flows around central AGN are stable over a few
million years, whereas we find that the mean jet power varies on time
scales of $\sim 10^7$~yr (the time between outbursts).

We wish to compare the energy in shocks to the energy in the X-ray
cavities for each outburst.
The total internal energy of
the cavities is 
roughly 3 times the mechanical energy ($P V$) required to inflate 
the cavities (McNamara \& Nulsen 2007), which is given in
Table~\ref{tab:cavities}.  We find total cavity internal energies of 
$8.6 \times 10^{56}$~erg and $7.8 \times 10^{55}$~erg for the 
outbursts that produced the intermediate and inner cavities,
respectively.  Thus, the total internal energy in cavities
is roughly 30\% of the shock energy for the previous outburst and
1.3 times the shock energy for the current outburst.
This is consistent with the current outburst being young, with the jets
actively inflating the inner cavities and driving the 1.5~kpc shock,
whereas for the previous outburst the shock has detached from the
cavities, which rise buoyantly and lose energy as they age.
We note that while our point explosion shock model is only
approximate, so that the shock energies are somewhat uncertain, the
relative sense of the energies is correct. Therefore, the larger
fraction of total energy in cavities in the current outburst (as
compared to the previous one) is a robust result.


For the outermost pair of cavities, the total internal energy in the
cavities is  
$\sim 9.6 \times 10^{56}$~erg, on the order of the total energy of the
intermediate cavities.  
However, 80\% of this energy is contained in the northeastern cavity.
While it is in principle possible for the mechanical energy to differ
between paired cavities from the same outburst,
the measurements for the southwestern cavity are uncertain since
it is only marginally detected.  Furthermore, such a large difference is not
seen in the inner and intermediate cavity pairs.
As discussed in \S~\ref{sec:cavities}, the outer cavities may be in
the process of breaking apart, and the outer northeastern cavity may
have been re-energized by the intermediate cavity, making the measured
mechanical energy (and hence the internal energy) uncertain.

We conclude that the lower
total energy of the most recent outburst, as compared to the previous
outburst, suggests that the outburst is ongoing, though this may
simply be an effect of the lower mean power. Although the
luminosity of the central source is low (see \S~\ref{sec:agn}), it may
be in a short term quiescent state or 
heavily obscured, and does not rule out an ongoing outburst.  AGN
luminosities are known to vary by several orders
of magnitude over very short time scales (\eg, Harris
\etal\ 2009).  Results from our 1D hydrodynamical simulations suggest
that the mean power over longer timescales ($\sim10^7$~yr)
can also vary  significantly between outbursts,
even in an otherwise relaxed system like NGC~5813.

\subsection{The Balance Between Heating and Cooling in the
  ICM}\label{sec:balance} 

As noted in \S~\ref{sec:energy}, the total mechanical energy output of
the AGN that is available to
heat the ICM is primarily in two forms: the internal energy of X-ray
cavities, which rise buoyantly after being inflated by jets from the
AGN, and the ``shock energy'', from shocks driven by the rapid inflation of
the cavities.  When and where the internal energy of the X-ray
cavities gets
transferred to the ICM is not well understood. In contrast, the
local heat input at the shock front can be calculated directly from the
Mach number.  Furthermore, shock heating has two desirable features.
First, the gas is more strongly heated in the
core where Mach numbers are larger, close to the central AGN, which is
the region of interest for regulating feedback between the ICM and the
central SMBH. Second, the heating is roughly isotropic (as opposed to
heating with jets or with the internal energy of the X-ray cavities).
We therefore consider the energy input due to shocks, which, as we
show in \S~\ref{sec:energy}, contain 40--80\% of the
total outburst energy in NGC~5813 (for a detailed discussion of
heating with AGN outburst shocks see David \etal\ 2001).

To balance radiative cooling with AGN feedback shocks, the average
outburst power must be on the order of or larger
than the rate of radiative cooling.
We estimated
the radiative cooling rate as the X-ray luminosity, which we obtained
by fitting the spectrum from the total emission 
within 170\arcsec\ (26.3~kpc) with an absorbed {\sc apec} model (using
a {\sc vapec} or two {\sc apec} model did not significantly change the
resulting luminosity).  The
derived X-ray luminosity of $L_X = 5.4 \times
10^{41}$~erg~s$^{-1}$ within 26~kpc implies a mean cooling time of
$t_{\rm cool} = 1.0$~Gyr (where we take the 
cooling time to be the time it would take to radiate away all of the
internal energy of the gas 
at the current luminosity).
For the 10~kpc shocks, we calculate the total shock energy to be
$\sim 3 \times 10^{57}$~erg (see \S~\ref{sec:energy}).  Therefore,
only 6 such outbursts
are needed per cooling time to provide enough energy to completely
offset radiative cooling with shocks alone within 26~kpc.
The time between outbursts given by both the buoyant rise times of
the cavities (Table~\ref{tab:cavities}) and, more reliably, from
hydrodynamical simulations of the shocks (Table~\ref{tab:shocks}) is
on the order of $10^7$~yr, allowing 100 shocks per cooling time, more
than is needed to provide the necessary energy. 

While results from the 10~kpc shocks indicate that there is sufficient
energy in shocks alone to offset radiative cooling, there are two
points that must be considered.  First, as discussed in
\S~\ref{sec:energy}, the observed shocks suggest that the
current and previous outburst shock energies differ in strength by
more than an order of magnitude (although the most recent outburst may
be ongoing).  Furthermore, only some fraction of the total
shock 
energy will go into heating the gas within the cooling radius, and for
weak shocks this fraction is relatively small ($\la 10$\%).
We therefore consider the {\it local} balance of shock heating and
radiative cooling for each shock.  
The primary effect of radiative cooling is to reduce the entropy of
the gas.  To offset cooling, the heating mechanism is
required to increase the entropy by at least this amount.
Shocks will also affect the kinetic, thermal, and potential energy of
the gas, but these effects are transient, and for heating it is the
change in entropy of the gas that is relevant.
Hence, the entropy increase $\Delta S$ caused by a weak shock
can offset a radiative heat loss of $\Delta Q \simeq T \, \Delta
S$, where $T$ is the gas temperature.
Expressed as a fraction of the gas thermal energy, the 
effective heat input from one shock
is therefore $(T \, \Delta S)/E = \Delta \ln (P / \rho^\gamma)$, where $E$ is
the thermal energy and $\gamma$ is the adiabatic index. 
For the 10~kpc shocks, the Mach numbers given in Table~\ref{tab:shocks}
imply a change in $\ln (P / \rho^\gamma)$ across
the shock front of $\sim 5\%$.  Therefore, to balance the
total entropy decrease of the gas 
about $1/0.05 = 20$ outbursts
are needed per {\it local} cooling time to completely offset cooling
with shock heating.
By the same argument, 10 outbursts per cooling time are needed to
replace the thermal
energy of the gas at the 1.5~kpc shock.  The cooling time
of the pre-shock gas just outside the 10~kpc shock edge is $9 \times
10^8$~yr, so an outburst interval of $10^7$~yr gives 90 shocks per
local cooling time.  For the 1.5~kpc shock, the pre-shock gas cooling
time is $2 \times 10^8$~yr, giving 20 shocks per cooling time.
Thus, we conclude that shock heating
alone is sufficient to offset radiative cooling of the gas within the
1.5~kpc and 10~kpc shock fronts, consistent with previous suggestions
that shocks can offset cooling close to central AGN (Nulsen \etal\ 2007).  
The X-ray cavities rise buoyantly, and release their internal energy
to heat the ICM gas at larger radii.
Although previous studies have found other systems where there is
enough total shock energy to offset radiative cooling
(\eg, M87 Forman \etal\ 2005, with M87 also showing concentric shock
fronts from multiple AGN outbursts; Hydra A Nulsen \etal\ 2005), here
we explicitly show that the fraction of shock energy that goes into
heating the gas (5--10\%) is sufficient to
balance radiative cooling locally at the shock fronts.  The outburst
interval is short enough for such shocks to offset cooling over much
longer timescales.
This demonstrates that AGN feedback can operate to heat the gas within
galaxies, as well as the more extended ICM in clusters, as required by
some galaxy formation models (\eg, Kormendy \etal\ 2009).


\section{Summary} \label{sec:summary}

We have presented results based on {\it Chandra}, {\it VLA}, {\it
  GMRT}, and {\it SOAR} observations 
of NGC~5813, the dominant member of a galaxy group.  The ICM shows
clear signatures from three distinct AGN outbursts in an otherwise
relaxed system (including two shocks with detectable temperature
  jumps), making this object uniquely well-suited to the study 
of AGN feedback.  We find the following:

\begin{enumerate}

\item
Three pairs of collinear cavities, where each pair is associated with a
distinct AGN outburst.  The inner two pairs are associated with
unambiguous shocks (with Mach numbers $M_i = 1.7$, $M_o = 1.5$), with
clear temperature rises, that can be directly
detected from the X-ray data.  
The outermost cavity pair also has an associated surface brightness
edge.  The properties of this edge are consistent with an old
shock associated with the outermost cavities, although the current
data do not rule
out other interpretations (\eg, a transition region from the
galactic atmosphere to the extended group atmosphere). 
The locations of the cavities and the
shocks indicate an outburst interval of $\sim10^7$~yrs.

\item
Diffuse radio emission, filling the inner cavities at 1.36~GHz and
235~MHz and the intermediate cavities at 235~MHz.  Radio emission from
the outer cavities is not detected.  This reflects the greater age of
the relativistic particles in more distant cavities.

\item
A cool trail of gas that has been buoyantly lifted by the intermediate
cavities.  The gas is co-spatial with filaments seen in H$\alpha$ observations.

\item
The mean power of the current outburst is six times less
than that of the previous outburst, indicating that the
average jet power can vary significantly between outbursts, even in
near ``steady state'' AGN feedback with regular outbursts in an
otherwise dynamically relaxed system.

\item
The total heat energy input from shocks alone is sufficient to balance
radiative cooling locally.
This heating takes place in the
core, close to the central AGN, which is the region of interest for regulating
feedback between the ICM and the central SMBH.  Thus, for the
first time, we explicitly show a system where shock heating alone can
locally balance radiative cooling and regulate AGN feedback.

\end{enumerate}


\acknowledgments
The financial support for this
work was partially provided for by the Chandra X-ray Center through
NASA contract NAS8-03060, and the Smithsonian Institution.
We thank the staff of the {\it GMRT} for their help during the
observations.  {\it GMRT} is run by the National Centre for Radio
Astrophysics of the Tata Institute of Fundamental Research.
The {\it SOAR} Telescope is a joint project of Conselho Nacional des
Pesquisas Cientficas e Tecnologicas CNPq-Brazil, The University of
North Carolina Chapel Hill, Michigan State University, and the
National Optical Astronomy Observatory.
AS and NW are supported by the National Aeronautics and Space Administration
through Chandra/Einstein Postdoctoral Fellowship Award Numbers PF9-00070 and
PF8-90056 issued by the {\it Chandra} X-ray Observatory Center, which
is operated by 
the Smithsonian Astrophysical Observatory for and on behalf of the National
Aeronautics and Space Administration under contract NAS8-03060.

\clearpage

\begin{deluxetable}{ccccccccc}
\tablewidth{0pt}
\tablecaption{Summary of the Radio Observations \label{tab:radio}}
\tablehead{
\colhead{Radio} & 
\colhead{Project} & 
\colhead{Observation}  & 
\colhead{Array} & 
\colhead{Frequency} & 
\colhead{Bandwidth} & 
\colhead{Integration}  & 
\colhead{HPBW, PA}  &   
\colhead{rms}  \\
\colhead{telescope} &
\colhead{}     &  
\colhead{date}  &
\colhead {}  & 
\colhead{(MHz)} &  
\colhead{(MHz)} &  
\colhead{time (min)}   &
\colhead{($^{\prime \prime} \times^{\prime \prime}$, $^{\circ}$)} &
\colhead{($\mu$Jy b$^{-1}$)}
}
\startdata
GMRT & 14SGA01 & Aug 2008 & full & 235  & 8 & 100 & $16.5 \times 15.1, 74$& 300 \\
VLA  & AF0188  & Apr 1990 &  A   & 1490 & 50 & 45 & $1.3 \times 1.2, 28$ &22 \\
VLA  & AW0202  & Jan 1988 &  B   & 1360 & 50 & 30 & $4.9 \times 4.7, -84$ &20 \\
VLA  & AC0488  & Sept 1997 & C   & 1360 & 50 & 6  & $27.4 \times 16.8, 53$& 30 \\
VLA  & AW0112  & Jun 1984  & C   & 4860 & 50 & 8  & $5.0 \times 4.5, -4$ &25 \\
\enddata
\end{deluxetable}

\begin{deluxetable}{lcccccccccc}
\tablewidth{0pt}
\tablecaption{Properties of the Shocks \label{tab:shocks}}
\tablehead{
\colhead{ID}&
\colhead{$r$\tablenotemark{a}}&
\colhead{$\Delta \rho$\tablenotemark{b}}&
\colhead{$M$\tablenotemark{c}}&
\colhead{$t_{\rm age,model}$\tablenotemark{d}}&
\colhead{$E_{\rm sh,model}$\tablenotemark{e}}&
\colhead{$W_{\rm model}$\tablenotemark{f}}&
\colhead{$M_{\rm acc}$\tablenotemark{g}}&
\colhead{$t_{\rm age,est}$\tablenotemark{h}}&
\colhead{$E_{\rm sh,est}$\tablenotemark{i}}&
\colhead{$W_{\rm est}$\tablenotemark{j}}\\
\colhead{}&
\colhead{(kpc)}&
\colhead{}&
\colhead{}&
\colhead{($10^7$ yr)}&
\colhead{($10^{57}$ erg)}&
\colhead{($10^{42}$ erg s$^{-1}$)}&
\colhead{($10^3 M_\odot$)}&
\colhead{($10^7$ yr)}&
\colhead{($10^{57}$ erg)}&
\colhead{($10^{42}$ erg s$^{-1}$)}
}
\startdata
Inner, SE&1.4&$1.97^{+0.12}_{-0.12}$&1.71&0.3&0.06&0.6&0.3&0.2&0.2&2.7\\
Middle, SE&9.9&$1.69^{+0.07}_{-0.07}$&1.48&1.4&2.2&5.0&8.2&1.6&3.0&5.9\\
Middle, NW&11.5&$1.75^{+0.04}_{-0.03}$&1.53&1.3&3.3&8.0&22.4&1.6&3.0&5.9\\
\enddata
\tablenotetext{a}{Distance from the AGN to the shock front.}
\tablenotetext{b}{Density jump at shock front.}
\tablenotetext{c}{Mach number.}
\tablenotetext{d}{Shock age, from the hydrodynamical model.  The point
explosion approximation gives large Mach numbers at early times, so
the shock ages are underestimated.}
\tablenotetext{e}{Shock energy, from the hydrodynamical model.}
\tablenotetext{f}{Mean shock power, from the hydrodynamical model.}
\tablenotetext{g}{Minimum accreted mass needed to power the model outburst
  (assuming 100\% efficiency).}
\tablenotetext{h}{Shock age, estimated as the travel time from the
  current position to the center point of the elliptical shock edge.
  Since the current Mach number is used to give a constant velocity,
  the shock ages are overestimated.}
\tablenotetext{i}{Shock energy, estimated using Equation~\ref{eq:eshock}.}
\tablenotetext{j}{Estimated mean shock power.}
\end{deluxetable}

\begin{deluxetable}{lccccc}
\tablewidth{0pt}
\tablecaption{Properties of the X-ray Cavities \label{tab:cavities}}
\tablehead{
\colhead{ID}&
\colhead{$a$\tablenotemark{a}}&
\colhead{$b$\tablenotemark{b}}&
\colhead{$r$\tablenotemark{c}}&
\colhead{$t_{\rm rise}$\tablenotemark{d}}&
\colhead{$E_{\rm mech}$\tablenotemark{e}}\\
\colhead{}&
\colhead{(kpc)}&
\colhead{(kpc)}&
\colhead{(kpc)}&
\colhead{($10^7$ yr)}&
\colhead{($10^{55}$ erg)}
}
\startdata
Inner, SW&0.95&0.95&1.3&0.6\tablenotemark{f}&1.1\\
Inner, NE&1.03&0.93&1.4&0.7\tablenotemark{f}&1.5\\
Middle, SW&3.9&3.9&7.7&3.6&15.3\\
Middle-1\tablenotemark{g}, NE&2.9&2.2&4.9&2.3&9.3\\
Middle-2\tablenotemark{g}, NE&2.8&2.4&9.3&4.4&4.1\\
Outer\tablenotemark{h}, SW&5.2&3.0&22.2&10.4&6.0\\
Outer, NE&8.0&4.4&18.0&8.5&26.0\\
\enddata
\tablenotetext{a}{Semi-major axis.}
\tablenotetext{b}{Semi-minor axis.}
\tablenotetext{c}{Distance from central AGN.}
\tablenotetext{d}{Lower limit on the bubble rise time, assuming that
  each bubble rises at half the sound speed for a 0.65~keV gas.}
\tablenotetext{e}{$P V$ mechanical energy required to inflate the cavity.}
\tablenotetext{f}{The cavity size is on the order of the distance to
  the AGN, and these cavities are likely still being or have only recently
  been inflated by the AGN, so the computed rise time is not a
  reliable estimate of the cavity age.}
\tablenotetext{g}{Part of a ``split'' or ``Russian doll'' cavity.}
\tablenotetext{h}{Cavity is only marginally detected, tabulated
  properties may not be reliable.}
\end{deluxetable}

\clearpage

\end{document}